\documentclass{aa}  

\usepackage{color}
\usepackage{graphicx}
\usepackage{txfonts}
\usepackage{hyperref}
\hypersetup{colorlinks=true,allcolors=[rgb]{0,0,0.5}}

\makeatletter
\renewcommand*\aa@pageof{, page \thepage{} of \pageref*{LastPage}}
\makeatother

\begin{document}

   \title{Spectral and angular differential imaging with SPHERE/IFS}

   \subtitle{Assessing the performance of various PCA-based approaches to PSF subtraction }

   \author{S. Kiefer \inst{1, 2}
          \and
          A. J. Bohn \inst{3}
          \and
          S. P. Quanz \inst{1}
          \and
          M. Kenworthy \inst{3}
          \and
          T. Stolker \inst{3, 1}
          }

   \institute{Institute for Particle Physics and Astrophysics, 
              ETH Zurich, Wolfgang-Pauli-Strasse 27, 8093 Zurich, Switzerland\\
              \email{kiefersv@student.ethz.ch}
         \and
             Visiting student at Leiden Observatory, University of Leiden         \and
             Leiden Observatory, University of Leiden, Niels Bohrweg 2, 2333 CA Leiden, The Netherlands
             }

   \date{Received ...; accepted ...}

  \abstract
   {Direct imaging of exoplanets is a challenging task that requires state-of-the-art instrumentation and advanced image-processing techniques. Differential imaging techniques have proven useful for the detection of exoplanet companions around stars. Angular differential imaging (ADI) and spectral differential imaging (SDI) are commonly used for direct detection 
and characterisation of young, Jovian exoplanets in datasets obtained with the SPHERE/IFS instrument.}
   {We compare the performance of ADI, SDI, and three combinations of ADI and SDI to find which technique achieves the highest signal-to-noise ratio (S/N), and we analyse their performance as functions of integration time, field rotation, and wavelength range.}
   {We analyse SPHERE/IFS observations of three known exoplanets, namely Beta~Pictoris~b, 51~Eridani~b, and HR~8799~e, with five differential imaging techniques. We split the datasets into subsets to vary each parameter before the data are processed with each technique. The differential imaging techniques are applied using principal component analysis (PCA).}
   {The tests show that a combination of SDI and ADI consistently achieves better results than ADI alone, and using SDI and ADI simultaneously (combined differential imaging; CODI) achieved the best results. The integration time test shows that targets with a separation larger than 0.24 arcsec observed with an integration time of more than 10$^3$s were photon-noise limited. Field rotation shows a strong correlation with S/N for field rotations up to 1 full width at half maximum (FWHM), after which no significant increase in S/N with field rotation is observed. Wavelength range variation shows a general increase in S/N for broader wavelength ranges, 
but no clear correlation is seen.}
   {Spectral information is essential to boost S/N compared to regular ADI. Our results suggest that CODI should be the preferred processing technique to search for new exoplanets with SPHERE/IFS. To optimise direct-imaging observations, the field rotation should exceed 1 FWHM to detect exoplanets at small separations.}


   \maketitle
%

\section{Introduction}
    Even though more than 4000 exoplanets have been discovered, only less 
than one percent of these have been directly imaged. Nevertheless, high-contrast imaging surveys like SHINE \citep{vigan_sphere_2020}, the GPI exoplanet survey \citep{nielsen_gemini_2013, nielsen_gemini_2019}, and the MMT-Clio survey \citep{heinze_constraints_2010} have assessed the demographics of giant exoplanets and brown dwarfs. To observe exoplanets with high-contrast imaging, state-of-the-art technology is required to overcome the 
two main challenges: high angular resolution and high contrast.
    
    To achieve high angular resolution, large telescopes (such as the 8.2-m UT telescopes of the Very Large Telescope (VLT) observatory) are required because of the Rayleigh criterion.  To achieve high contrasts, the light of the star must be suppressed, as the diffraction halos of the target stars are typically several orders of magnitude brighter than an accompanying exoplanet. At the VLT, the Spectro-Polarimetric High-contrast Exoplanet Research \citep[SPHERE;][]{beuzit_sphere_2019} instrument is equipped with coronagraphs. Its extreme adaptive optics (AO) system helps to reduce speckle noise introduced by atmospheric turbulence and to further increase the final contrast. To obtain spectra, SPHERE is equipped with an integral field spectrograph (IFS) which can measure low-resolution spectra \citep{claudi_sphere_2008}. The wavelength range can be set to 0.95 - 1.35 $\mu$m with 
a resolution of R $\approx$ 50, or it can be set to 0.95 - 1.68 $\mu$m with a spectral resolution of R $\approx$ 30. In both cases, there are 39 wavelength channels split over the wavelength range \citep{wahhaj_sphere_2019}. The field of view (FOV) of SPHERE/IFS is approximately $1.73''\times 1.73''$.
        
    State-of-the-art technology alone is not sufficient -- advanced data-reduction algorithms are necessary to increase the achievable contrast performance in data post-processing. Angular differential imaging (ADI) as described by \citet{marois_angular_2006} (see section \ref{ssec:ADIandSDI}) lays the foundation for most advanced techniques used today. Examples of such algorithms are LOCI \citep{lafreniere_new_2007}, TLOCI \citep{marois_tloci_2013,marois_gpi_2014}, MLOCI \citep{wahhaj_improving_2015}, ANDROMEDA \citet{mugnier_optimal_2009, cantalloube_direct_2015}, PACO \citep{flasseur_exoplanet_2018}, and TRAP \citet{samland_trap_2020}. Lastly, there are attempts to use machine-learning techniques to improve exoplanet detections in high-contrast imaging \citep{gonzalez_supervised_2018, gebhard_physically_2020}.
    
    \citet{amara_pynpoint_2012} and \citet{stolker_pynpoint_2019} developed the Python package PynPoint for processing and analysis of high-contrast imaging data taken by, for example, SPHERE or NACO \citep[NACO;][]{lenzen_naos-conica_2003, rousset_naos--first_2003}. PynPoint uses principal component analysis \citep[PCA;][]{amara_pynpoint_2012,soummer_detection_2012} and contains many functionalities to further analyse and process data (e.g. calculate contrast limits); a full list can be found in the official PynPoint documentation\footnote{Available at \url{https://pynpoint.readthedocs.io/}}. However, the basic ADI assumption of a quasi-static residual stellar point spread function (PSF) only holds partially because of changes in the AO performance, mostly due to non-common path aberrations (NCPA). This leads to an imperfect PSF model and subtraction can create residuals that resemble a planet. Spectral differential imaging \citep[SDI;][]{racine_speckle_1999} on the other hand uses images taken simultaneously and therefore with the same NCPAs (see section \ref{ssec:ADIandSDI}). Recently, the PACO-ASDI algorithm \citep{flasseur_paco_2020} was developed, which uses IFS data to add SDI (in combination with ADI) to PACO. These latter authors found that adding SDI to ADI results in improved sensitivity of their detection maps. 
    
     Other pipelines similar to PynPoint have been developed for an efficient analysis of high-contrast imaging data. Examples of such pipelines are the Vortex Image Processing Package \citep[VIP;][]{gonzalez_vip_2017}, the SPHERE speckle calibration tool \citep[SpeCal;][]{galicher_astrometric_2018}, and the automated data-processing architecture for the Gemini Planet Imager Exoplanet Survey \citep{wang_automated_2018}.
    
    Spectral differential imaging and combinations of SDI and ADI for exoplanet imaging surveys are 
already in use. For example, \citet{biller_gemininici_2013}, \citet{nielsen_gemini_2013}, and \citet{wahhaj_gemini_2013} conducted a survey using SDI in combination with ADI. For their SDI, they took two images simultaneously at two different wavelengths. They set the first wavelength channel 
at a methane absorption line (1.652 $\mu$m) so that T-type substellar objects would not appear in the image. The second wavelength was chosen outside the methane band (1.578 $\mu$m). The first image was then used as a PSF model to correct the second one. \citet{samland_spectral_2017} and \citet{zurlo_first_2016} applied different combinations of ADI and SDI to detect and characterise the exoplanets of 51 Eridani and HR 8799, respectively, using 39 wavelength channels to achieve a more accurate PSF model. \citet{biller_gemininici_2013}, \citet{nielsen_gemini_2013}, \citet{wahhaj_gemini_2013}, \citet{samland_spectral_2017}, and \citet{zurlo_first_2016} all used ADI and SDI one after the other. In contrast, the algorithm of \citet{flasseur_paco_2020} combines ADI simultaneously with SDI. This is possible, as both techniques exploit the diversity of sky rotation of a potential exoplanet companion (see section \ref{ssec:ADIandSDI}). \citet{christiaens_separating_2019} compared the performance of applying ADI followed by SDI with the output when alternating the order of both processing methods (i.e. SDI followed by ADI). These authors used different post-processing techniques on PDS 70, a K7-type star that hosts two gas giant planets \citep{keppler_discovery_2018, haffert_two_2019} which are embedded in its transitional disc. Injecting fake planets allowed them to study which differential imaging technique best recovers specific features.
    
    In this paper, we looked into different combinations of ADI and SDI to determine which yields the highest S/N detection of directly imaged planets. Even though various ADI/SDI combinations have been developed \citep[e.g.][]{zurlo_first_2016, samland_spectral_2017, flasseur_exoplanet_2018}, they have not yet been extensively compared to each other. To do so, we added IFS support to PynPoint allowing wavelength-dependent differential imaging techniques as described in Section \ref{sec:DifferntialImaging}. In Section \ref{sec:Results}, we use this implementation to reanalyse 
data on Beta Pictoris, 51 Eridani, and HR 8799 and to calculate the S/N of their exoplanets. We then discuss the results in Section \ref{sec:Discussion} in order to decipher which combination of ADI and SDI achieves the highest S/N. In Sections \ref{sec:Results} and \ref{sec:Discussion}, we determine the  dependence of the differential imaging techniques on field rotation, wavelength range, and total integration time. In Section \ref{sec:summary}, we summarise our findings.

\section{Differential Imaging}
\label{sec:DifferntialImaging}

    \subsection{ADI and SDI}
    \label{ssec:ADIandSDI}
        The primary goal is to understand how ADI and SDI can be effectively combined to create a PSF model that provides the best possible contrast performance. To use ADI, raw images are taken with the instrument derotator turned off. This causes the FOV to rotate around the target star during observations. The stellar PSF is, to first order, quasi-static but a potential exoplanet rotates around the stellar position during the observation. 
The images can be used to build a model of the stellar PSF with negligible contribution from the companion, which can then be subtracted from each 
individual image. After de-rotating and stacking the images, a possible companion should become visible. In this paper, ADI is used in combination 
with PCA to optimise planet detection. From the input images, an orthogonal basis is calculated where the components are sorted with decreasing variance. The elements of this basis set are called principal components (PCs). A companion should not influence the variance significantly. Thus, the first few PCs can be used to create a PSF model while minimising the influence of planetary signals. For more information on PCA the reader is referred to \citet{jee_principal_2007}. To perform ADI together with PCA, we used PynPoint as described by \citet{amara_pynpoint_2012} and \citet{stolker_pynpoint_2019}.
        
        To perform SDI, PynPoint was extended to support SPHERE/IFS data. Like ADI, our SDI implementation uses displacements of a possible companion to model the stellar PSF while minimising the contribution of a potential exoplanet. 
                
   \begin{figure*}
   \centering
   \includegraphics[width=0.8\textwidth]{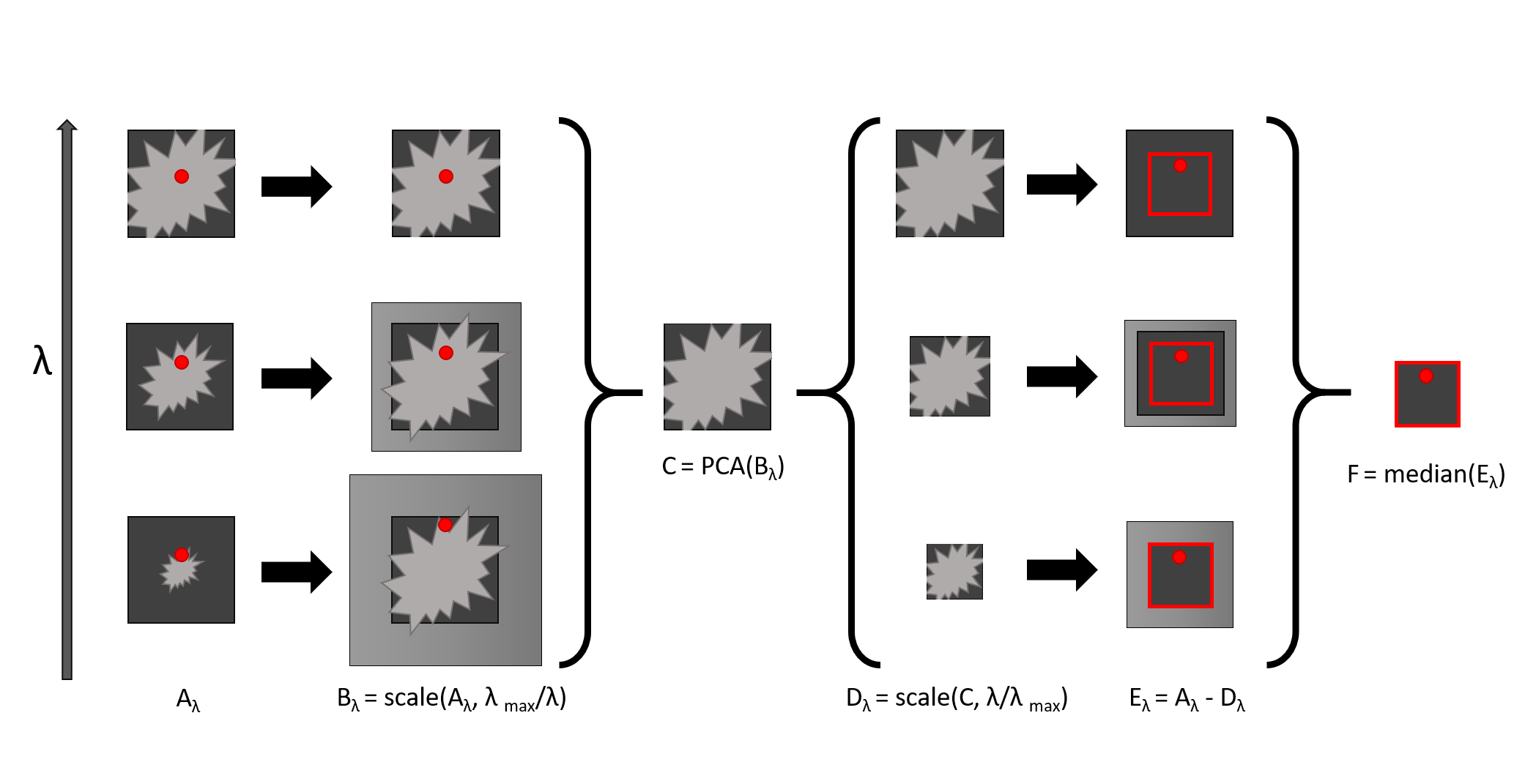}
      \caption{Schematic of SDI (including the FOV loss through scaling). 
First, the images (A$_\lambda$) are scaled inversely proportional to their wavelength (B$_\lambda$). A stellar PSF model (C) is then created using 
PCA analysis. Afterwards the stellar PSF model (C) is rescaled for each wavelength channel proportional to its wavelength (D$_\lambda$) to have the same scaling as the original images (A$_\lambda$). Therefore, the outer 
region of the small wavelength images is not covered by the scaled stellar PSF models (D$_\lambda$). This reduces the FOV by a factor R = $\lambda_{min}$/$\lambda_{max}$. The reduced FOV is marked by a red border, the 
companion is marked as a red dot.
              }
         \label{fig:SDIscaling}
   \end{figure*}
   
        Because of the diffraction caused by the telescope, the flux from a distant point source will be spread out into its PSF. The shape of the PSF is 
approximately the same for all wavelengths, but with different full width at 
half maximum (FWHM) according to the Rayleigh criterion:
        \begin{equation}
                \beta (\lambda) \propto \frac{\lambda}{\mathcal{D}}\;,
        \end{equation} 
        where $\mathcal{D}$ is the telescope diameter and $\lambda$ the observed 
wavelength. Scaling images that are obtained at different wavelengths with a factor $S \propto 1 / \lambda$ corrects for the wavelength dependence 
of the PSF size. At the same time, the scaling causes a radial shift of a 
possible companion. The conditions are equivalent to ADI which uses a quasi-constant stellar PSF over time and azimuthal displacements of a possible companion. Therefore, a similar processing framework can be used for ADI and SDI.
        
        To scale the images and thus align the stellar PSF, the inverse of $\beta (\lambda)$ can be used. Because only the relative scaling between the images matters, the scaling factor can be defined as:
        \begin{align}
                \label{eq:scalingFactor}
                S(\lambda) &= \frac{\beta (\lambda_\mathrm{ref})}{\beta (\lambda)} = 
\frac{\lambda_\mathrm{ref}}{\lambda.}
        \end{align}
        We choose the reference wavelength $\lambda_{\mathrm{ref}}$ as the longest observed wavelength to prevent under-sampling of the residual star light caused by downscaling images. 
        \begin{equation}
            \lambda_\mathrm{ref} := max\{\lambda\}
        .\end{equation}
        Using this reference wavelength results in an up-scaling of all images with a shorter wavelength ($S (\lambda) \geq 1 , \forall \lambda$).
The scaling of the images leads to a reduced FOV as illustrated in Fig. \ref{fig:SDIscaling}. SPHERE/IFS has a FOV of $1.73 \times 1.73$ arcsec$^2$ and a spectral range of 0.95 - 1.35 $\mu$m (or 0.95 - 1.68 $\mu$m) \citep{wahhaj_sphere_2019}. Scaling the images with the scaling factor S($\lambda$) leads to a final FOV of 1.22 arcsec (or 1.00 arcsec). Everything 
outside this range is not covered by all wavelengths and therefore the stellar PSF cannot be modelled well. This FOV reduction does not occur if only ADI is applied.
        
        We did not select specific wavelength channels for SDI and used all the available images. This confers the advantage that no assumptions about a possible companion have to be made and the results can be compared between different observations. Other papers removed certain wavelengths to either reduce self-subtraction or exploit known absorption bands \citep[e.g.][]{thatte_very_2007, samland_spectral_2017}. If using the differential imaging techniques discussed in this paper to search for exoplanets, observation specific wavelength selection for SDI could increase the performance.
        
        \subsection{Displacements}
        The radial displacements $\Delta r$ of a companion at separation r are caused by the scaling of the images according to Eq. \ref{eq:scalingFactor}. The smaller the displacements, the more planet signal is present in the PSF model. Large radial displacements are therefore desirable. The total radial displacements $\Delta r$ between the shortest ($\lambda_\mathrm{min}$) and longest ($\lambda_\mathrm{max}$) observed wavelength at the separation of a companion $r_\mathrm{pl}$ can be calculated with the following equation:
        \begin{align}
            \label{eq:DeltaR}
            \Delta r &
            = \frac{\lambda_\mathrm{max}}{\lambda_\mathrm{min}} r_\mathrm{pl} - r_\mathrm{pl}
            = r_\mathrm{pl}\frac{\Delta \lambda}{\lambda_\mathrm{min}},
        \end{align}
        where $\Delta \lambda = \lambda_\mathrm{max} - \lambda_\mathrm{min}$ is the observed wavelength range. 
        
        The azimuthal displacements $\Delta \mathrm{rot}$ of a companion used in 
ADI are due to the field rotation during the observation. Because the companion shifts its position over the circular segment with an angle of $\Delta \alpha$ and radius $r_\mathrm{pl}$, $\Delta \mathrm{rot}$ can be calculated with the following equation:
        \begin{align}
        \label{eq:DeltaRot}
            \Delta \mathrm{rot} &
            = r_\mathrm{pl} \Delta \alpha.
        \end{align}
        In our analysis, we did not set a required minimum displacement. We analysed a range of different field rotations and wavelength ranges (see Section \ref{sec:diff_imag_tests}) to find their influence on the performance 
of differential imaging techniques.

    \subsection{Differential imaging techniques}
    \label{ssec:ADI_SDI_Combination}
    Both ADI and SDI use displacements of possible companions to create a 
PSF model excluding most of the signal from the  companion. Therefore, these two techniques can be combined in three different ways: using first ADI then SDI; using first SDI then ADI; or using both techniques simultaneously. The implementation of all five techniques is described below.
        
        \begin{itemize}
                \item \textbf{ADI:} First, all images $I_{\lambda,t}$ within the dataset $\mathcal{H}=\{I_{\lambda,t}\,\forall \lambda, t\}$ are split into wavelength subsets $\mathcal{H}_{\Tilde{\lambda}} = \{I_{\lambda,t} \mid \lambda = \Tilde{\lambda}\}$. Afterwards, using only the images of the subset $\mathcal{H}_{\Tilde{\lambda}}$, a PSF model $\mathcal{P}_{\Tilde{\lambda},t}$ is created using PCA with $n \in \mathbb{N}$ PCs\footnote{\label{fn:n_expl}$n$ has the same value for all subsets.}. Each image $I_{\Tilde{\lambda,t}}$ is then subtracted by the PSF model $\mathcal{P}_{\Tilde{\lambda},t}$. Lastly, all images $I_{\lambda,t}$ are derotated and stacked using the median in order to create the final image $I_{ADI}$.
                
                \item \textbf{SDI:} First, all images $I_{\lambda,t}$ are scaled according to $\lambda$ using a scaling factor $S(\lambda) = \lambda_{max} / \lambda$. Afterwards, the dataset $\mathcal{H}=\{I_{\lambda,t}\,\forall \lambda, t\}$ is split into time subsets $\mathcal{H}_{\Tilde{t}} = \{I_{\lambda,t} \mid t = \Tilde{t}\}$. Using PCA with $n$ PCs\textsuperscript{\ref{fn:n_expl}}, a PSF model $\mathcal{P}_{\lambda,\Tilde{t}}$ is created using only images from the subset $\mathcal{H}_{\Tilde{t}}$. Each image $I_{\lambda,\Tilde{t}}$ is then subtracted by the PSF model $\mathcal{P}_{\lambda,\Tilde{t}}$. Lastly, all images $I_{\lambda,t}$ are scaled back (using a factor $1/S(\lambda)$) before they are derotated and stacked using the median  in order to create the final image $I_{SDI}$.
                
                \item \textbf{CODI (combined differential imaging):} First, all images $I_{\lambda,t}$ are scaled according to $\lambda$ using a scaling factor $S(\lambda) = \lambda_{max} / \lambda$. Afterwards, a PSF model $\mathcal{P}_{\lambda,t}$ is created using PCA with $n$ PCs\footnote{We note that in contrast to all other differential imaging techniques, CODI uses only one PCA using $n$ PCs.} The PSF model $\mathcal{P}_{\lambda,\Tilde{t}}$ is 
then subtracted from each image $I_{\lambda,\Tilde{t}}$. Lastly, all images $I_{\lambda,t}$ are scaled back (using a factor $1/S(\lambda)$) before 
they are derotated and stacked using the median in order to create the final image $I_{CODI}$.
                
                \item \textbf{SADI:} First SDI is performed as described above but without derotating and stacking after the reduction. Afterwards it performs ADI as described above to create the final image $I_{SADI}$. Both steps use $n$/2 PCs where $n$ is an even number.
                
                \item \textbf{ASDI:} First ADI is performed as described above but without derotating and stacking after the reduction. Afterwards it performs SDI as described above to create the final image $I_{ASDI}$. Both steps use $n$/2 PCs where $n$ is an even number.
        \end{itemize}
        
        The differential imaging techniques and the framework to process SPHERE/IFS data are available within the PynPoint library. Furthermore, all images were pre-processed using EsoReflex \citep{freudling_automated_2013}. This includes flat-field corrections, dark-frame subtraction, and wavelength calibration. For the PCA, we use the entire subset to create the PCA basis for the PSF model. The first few elements of this basis are than fitted to each image of the subset individually to create the PSF models. 

        Theoretically, every SDI step can be followed by an ADI step, or vice versa, therefore creating many more possible combinations. We focused on techniques using a maximum of one ADI step and one SDI step, one after the other 
(namely SADI and ASDI). Furthermore, each combination also allows different numbers of PCs to be used in each ADI or SDI step. We focused on using 
the same number of PCs in all ADI and SDI steps to be able to test a large range of PCs.
        
        As ADI splits the dataset into its wavelength channels, it will conduct a PCA for each wavelength. For example, if ADI is performed on SPHERE/IFS 
data, a total of 39 PCAs will be conducted. As SDI splits the dataset into 
time components, it will conduct a PCA for each time instance of the dataset. If SDI is applied to a dataset with 100 time instances (each containing 39 wavelength frames), this leads to 100 PCAs. CODI does not split the dataset at all, and therefore only uses 1 PCA. To compare the results between different differential imaging techniques, we calculated the model completeness $\Omega$ by dividing the number of PCs used by the total number of PCs available:
        \begin{equation}
            \label{eq:Omega}
            \Omega = \frac{\mathrm{PC}_{\mathrm{used}}}{\mathrm{PC}_{\mathrm{max}}}
        .\end{equation}
        Therefore, the residuals are zero if a model completeness of 1 is used because the image is fitted perfectly.
        
        For SDI, SADI, and ASDI, the number of wavelength channels is the limiting factor and therefore PC$_{max} = 39$. For ADI, PC$_{max} = \mathrm{N}_\mathrm{tot, \lambda}$, where N$_\mathrm{tot, \lambda}$ is the number of images per wavelength taken during an observation. For CODI, PC$_{max} = 
39 \times \mathrm{N}_\mathrm{tot, \lambda}$ because it uses all images at 
once for a single PCA. A summary can be found in Table \ref{tab:PCmax}.

    \subsection{Differential imaging tests}
    \label{sec:diff_imag_tests}
    We test for dependence on total integration time, observed rotation, and/or observed wavelength range with the following:
    \begin{itemize}
        \item \textbf{Integration time:} We split each dataset into subsets with different total integration times. Each dataset starts at the first image and contains a number of subsequent images until the given integration time of the subset is reached. Afterwards, each differential imaging technique is applied separately to each subset. The same PC number was 
chosen for all subsets. We compare the results to a slope proportional to 
$\sqrt{t}$. If the S/N values follow this curve, they behave in a photon-noise-limited manner.
        
        \item \textbf{Field rotation:} We split each dataset into subsets 
with an equal number of images $M$. In the first subset, we select the first M images ($1, 2, \dots, M$); in the second, every second image is selected ($1, 3, \dots, 2 M$); in the third set every third image is selected ($1, 4, \dots, 3M$), and so forth. Therefore, all subsets have the same integration time but a different total field rotation $\Delta \alpha$. Because small rotations lead to a PSF overlap of a potential exoplanet, it is expected that small rotations lead to lower S/N. To achieve comparable 
results, we related the angular rotation to displacements expressed in FWHM using Eq. \ref{eq:DeltaRot} with $\Delta \mathrm{rot} = \mathrm{FWHM}$.
        
        \item \textbf{Wavelength range:} We split each dataset into subsets with an equal number of wavelength frames from all images. In the first subset, we select the first F frames from each image ($1, 2, \dots, F$); in the second, every second frame of each image is selected ($1, 3, \dots, 2 F$); in the third set every third frame of each image is selected ($1, 4, \dots, 3F$), and so forth. Similar to the rotational overlap, radial overlap (after scaling the images) can lead to a lower S/N. To achieve comparable results, we related the radial offsets to displacements expressed in FWHM using Eq. \ref{eq:DeltaR} with $\Delta r = \mathrm{FWHM}$.
    \end{itemize}

\section{Results}
\label{sec:Results}

    \begin{table*}
    \caption{All observations were performed with SPHERE/IFS (PI: J. Beuzit). Here, DIT stands for detector integration time and N$_\mathrm{tot, \lambda}$ is the total number of images taken per wavelength. Each image contains 39 wavelength frames equally spread over the wavelength range $\lambda$. Each dataset has a field rotation of $\Delta \alpha$. The atmospheric 
seeing is given as the approximate value for the atmospheric distortions during the whole observation.}             
    \label{table:TestingDataSets}      
        \centering 
                \begin{tabular}{c c c c c c c c}
                        \hline\hline
                         Object Name  & ESO ID        & Date       & DIT[s] x N$_\mathrm{tot, \lambda}$ & $\Delta$t[min] & $\Delta \alpha$[$^\circ$] & $\lambda$[$\mu$m] & Seeing['']  \\ \hline
                        Beta Pictoris & 096.C-0241(B) & 2015-11-30 & 4 x 760       & 50.7      
     & 41.1                      & 0.95-1.35         & $\approx$ 1 \\
                           HR 8799    & 095.C-0298(D) & 2015-09-28 & 16 x 256      & 68.3      
     & 23.8                      & 0.95-1.68         & $\approx$ 1 \\
                         51 Eridani   & 198.C-0209(J) & 2017-09-28 & 32 x 154      & 83.1      
     & 52.0                      & 0.95-1.68         & $<$ 0.8     \\ \hline\hline
                \end{tabular}
    \end{table*}
    
    \begin{table}
        \centering
        \caption{Maximum number of PCs, PC$_{max}$ , that can be selected for PCA. N$_\mathrm{tot, \lambda}$ is the total number of images taken per 
wavelength within the analysed dataset. The maximum number of SDIs is given by the number of wavelength channels observed (for SPHERE/IFS: 39). Because SADI and ASDI use SDI, they have the same maximum number of PCs.}
        \label{tab:PCmax}
        \begin{tabular}{c c c c c c}
            \hline\hline
             & ADI & SDI & CODI & SADI & ASDI \\  
             \hline
            $\mathrm{PC}_{max}$ & N$_\mathrm{tot, \lambda}$ & 39 & N$_\mathrm{tot, \lambda}$ $\times$ 39 & 39 & 39 \\
            \hline
        \end{tabular}
    \end{table}
    
    \begin{table}
    \caption{Number of principal components $n$ that overall achieve the 
highest S/N during the integration time test. The numbers of SADI, ASDI, and SDI were searched on a grid with $\Delta$PC = 1; ADI was searched with $\Delta$PC = 5; CODI was searched with $\Delta$PC = 10.}
    \label{tab:S/N}
    \centering
    \begin{tabular}{c c c c c c}
    \hline\hline
     & ADI & SDI & CODI & SADI & ASDI \\
    \hline
       Beta~Pictoris~b & 50 & 5 & 330 & 4 & 4 \\
       51~Eridani~b & 6 & 2 & 80 & 2 & 2 \\
    \hline
    \end{tabular}
    \end{table}
        
    \begin{figure}
    \centering
    \includegraphics[width=\hsize]{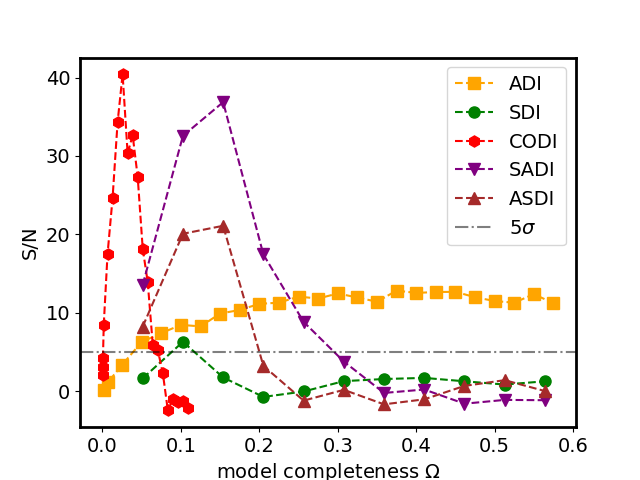}
    \includegraphics[width=\hsize]{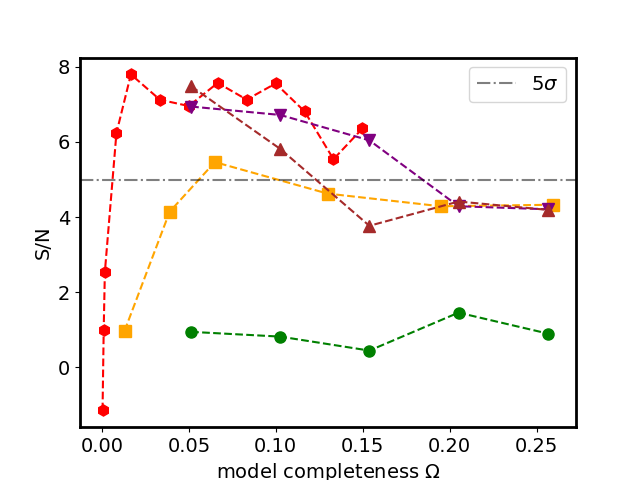}
    \includegraphics[width=\hsize]{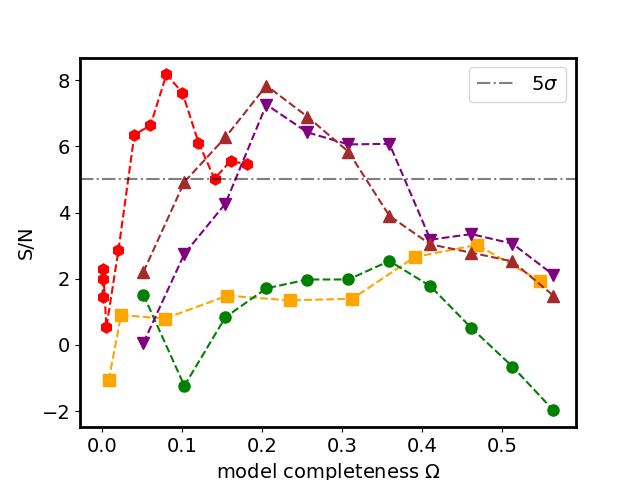}
    \caption{S/N as a function of model completeness using different differential imaging techniques. \textbf{Top: }Beta~Pictoris~b; \textbf{Middle:} 51~Eridani~b; \textbf{Bottom:} HR~8799~e.}
         \label{FigDetection}
   \end{figure}
    
    \begin{figure}
    \centering
    \includegraphics[width=\hsize]{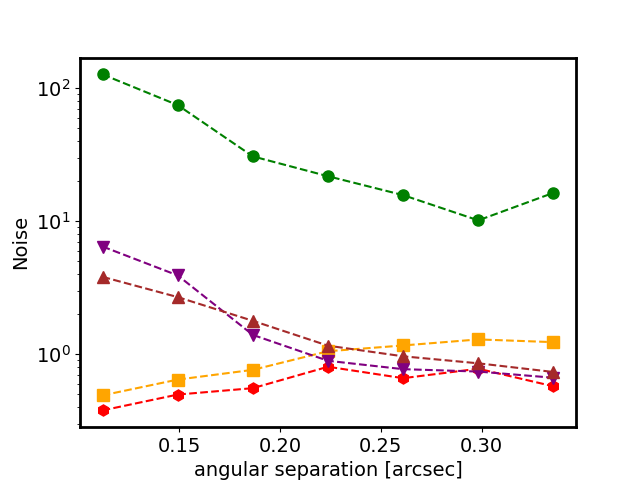}
    \includegraphics[width=\hsize]{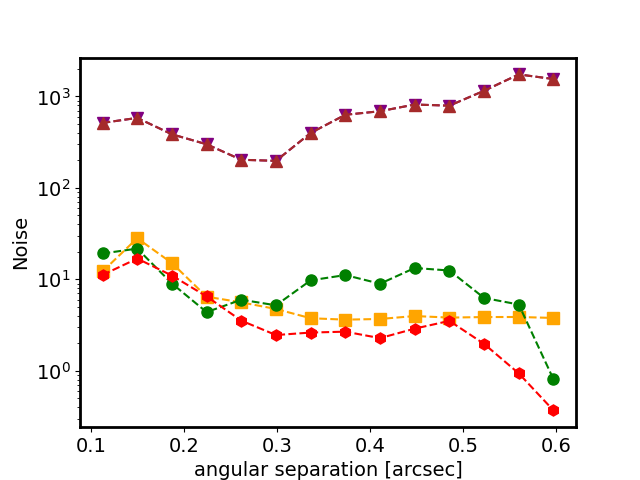}
    \includegraphics[width=\hsize]{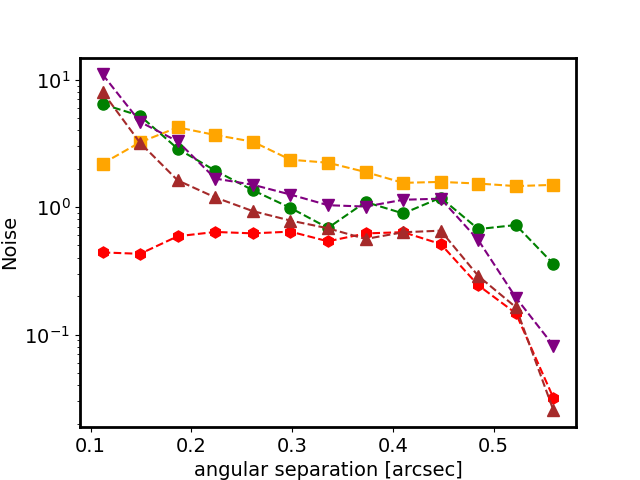}
    \caption{Noise as a function of angular separation using different differential imaging techniques. \textbf{Top: }Beta~Pictoris~b; \textbf{Middle:} 51~Eridani~b; \textbf{Bottom:} HR~8799~e. The colours and symbols are the same as in Fig. \ref{FigDetection}. The noise calculation is described in Section \ref{sec::sn_calc}.}
         \label{FigSNmap}
   \end{figure}
   
    \begin{figure}
    \centering
    \includegraphics[width=\hsize]{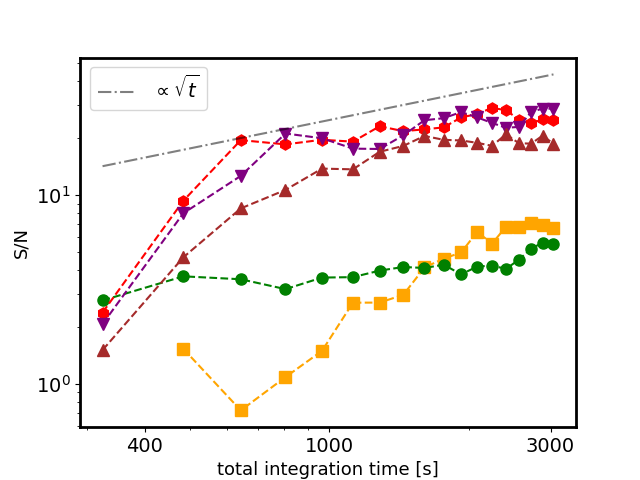}
    \includegraphics[width=\hsize]{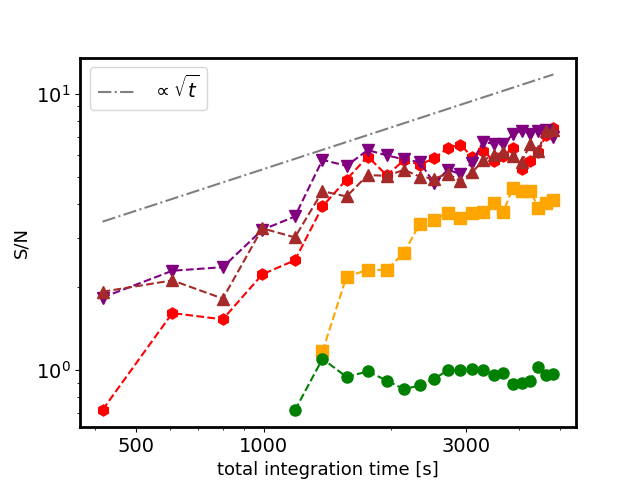}
    \caption{Dependence of S/N  on time after 
             processing all images using different differential 
             imaging techniques. \textbf{Top:} Beta~Pictoris~b; \textbf{Bottom:} 51~Eridani~b. The colours and symbols are the same as in Fig. \ref{FigDetection}. The grey line indicates the slope corresponding to photon-noise-limited behaviour.}
         \label{FigTime}
    \end{figure}
        
        \begin{figure}
    \centering
    \includegraphics[width=\hsize]{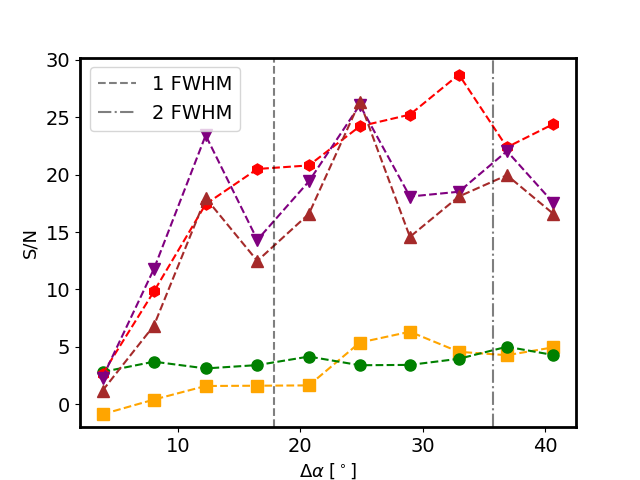}
    \includegraphics[width=\hsize]{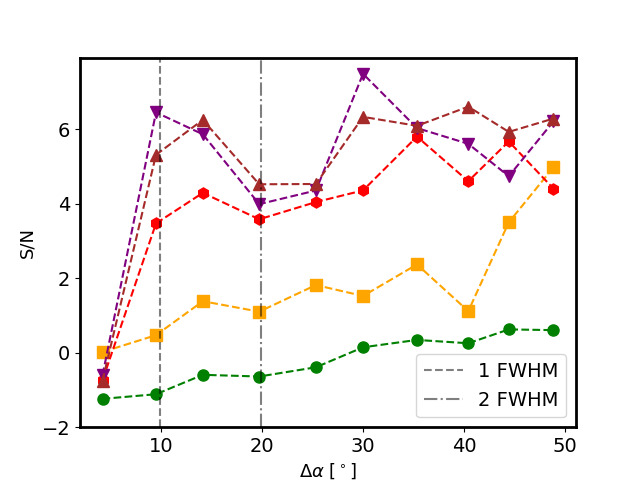}
    \caption{Dependence of S/N  on field rotation after 
             processing all images using different differential 
             imaging techniques. \textbf{Top:} Beta~Pictoris~b; \textbf{Bottom:} 51~Eridani~b. The colours and symbols are the same as in Fig. \ref{FigDetection}. The vertical grey lines indicate displacements of 1 and 2 FWHM according to Eq. \ref{eq:DeltaRot}.}
         \label{FigAngle}
    \end{figure}
                        
        \begin{figure}
    \centering
    \includegraphics[width=\hsize]{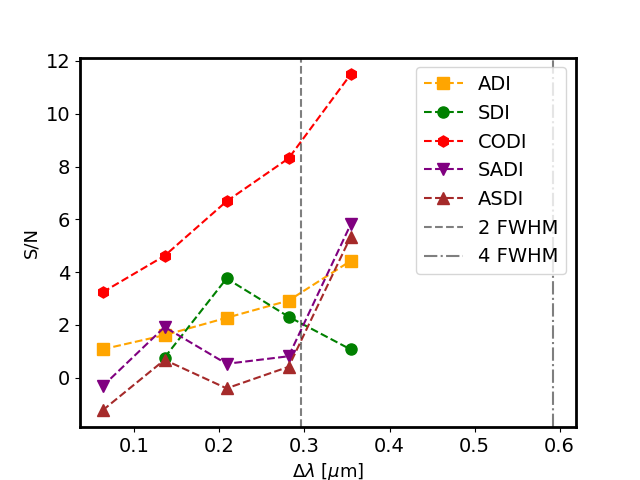}    \includegraphics[width=\hsize]{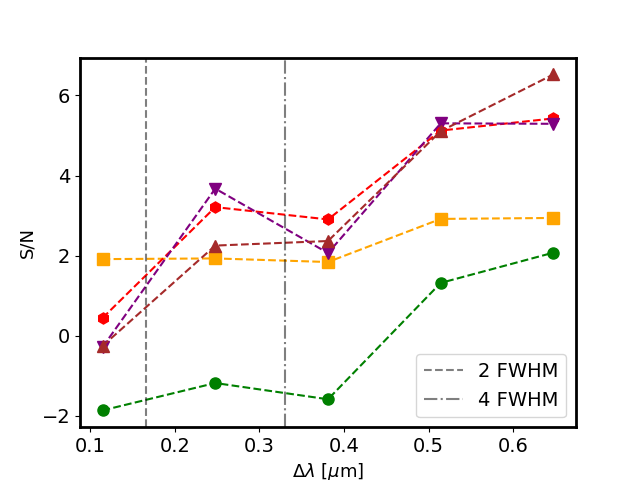}    \caption{Dependence of S/N  on wavelength range of Beta~Pictoris~b after              processing all images using different differential 
             imaging techniques. \textbf{Top:} Beta~Pictoris~b; \textbf{Bottom:} 51~Eridani~b. The colours and symbols are the same as in Fig. \ref{FigDetection}. The vertical grey lines indicate displacements of 2 and 4 FWHMs according to Eq. \ref{eq:DeltaR}.}
         \label{FigWave}
    \end{figure}
    
    \subsection{Datasets}
    \label{sec:Data_Sets}
    We used three stars with known exoplanets to compare the five different differential imaging techniques described in Section \ref{ssec:ADI_SDI_Combination}: Beta~Pictoris~b \citep{lagrange_giant_2010}, 51~Eridani~b \citep{macintosh_discovery_2015}, and HR~8799~e \citep{marois_images_2010}. The information on their observation can be found in Table \ref{table:TestingDataSets}.
    
    Among the three targets, Beta~Pictoris~b has the most favourable contrast. The exoplanet has a contrast of 10.4$\pm$0.5 mag in the $J$-band \citep{bonnefoy_near-infrared_2013}. In the selected dataset, Beta~Pictoris~b is at a separation of 0.242$\pm$0.0025 arcsec \citep{lagrange_post-conjunction_2019}. As a second target, we chose 51 Eridani. Its exoplanet 51~Eridani~b has a contrast of 15$\pm$1.6 mag in the $J$-band \citep{samland_spectral_2017} and is at a separation of 0.455$\pm$0.006 arcsec \citep{rosa_astrometric_2015}. The last test system we selected is HR 8799. From 
the four planets of the system, only HR~8799~`e' and `d' are within the FOV of SPHERE/IFS. After the scaling necessary for SDI (see Section \ref{ssec:ADIandSDI} and Fig. \ref{fig:SDIscaling}), only HR~8799~e remains within the reduced FOV. This exoplanet has a contrast of 13.01$\pm$0.21 mag in the 
$J$-band \citep{zurlo_first_2016} and is at a separation of 0.381$\pm$0.007 arcsec \citep{apai_high-cadence_2016}.

        \subsection{Calculating the S/N}
        \label{sec::sn_calc}
    Our goal was to quantify detections of exoplanets to compare the differential imaging techniques. We did not aim to assess the detection limitations of our techniques, nor did we aim to directly compare our detections with those of other studies. Therefore, we chose S/N calculations to compare the different differential imaging techniques.
        
        The signal used in the S/N was calculated using aperture photometry at the position of a companion candidate. All reference apertures for the noise calculation were chosen at the same angular separation as the signal aperture \citep{mawet_fundamental_2014}. The size of the apertures was chosen to be 1 FWHM of the PSF at the median wavelength observed ($\lambda = 
1.15$ for Beta Pictoris and $\lambda = 1.31$ for HR 8799 and 51 Eridani). To determine the position, a 2D Gaussian is fitted to the PSF of the  companion. To calculate the noise, we first calculate the sum of the pixel values within each reference aperture. The noise is then defined as the root mean square of the reference aperture values. The signal is calculated as 
the sum of the pixel values within the signal aperture. To reduce the influence of the planetary signal within the noise calculation, the apertures adjacent to the planetary signal were not used. The noise calculation is adjusted for small sample statistics \citep{mawet_fundamental_2014}.

    \subsection{Differential imaging tests}
    For comparability between differential imaging techniques, the number 
of PCs used is given in model completeness $\Omega$ (see Eq. \ref{eq:Omega}). The values for PC$_{max}$ can be seen in Table \ref{tab:PCmax}.

    The model completeness $\Omega$ was limited to 0.6 (0.25 for 51 Eridani) because all differential imaging techniques showed a clearly reduced S/N for larger model-completion values. The resulting S/N for Beta~Pictoris~b, 51~Eridani~b, and HR~8799~e depending on model completeness can be seen in Fig. \ref{FigDetection}. In all three targets, CODI achieves the highest S/N. SADI and ASDI achieve comparable results with the exception of SADI applied to Beta~Pictoris~b which results in a lower S/N than the other three techniques. Furthermore, the differential imaging techniques SADI, ASDI, and CODI always achieve higher S/N than ADI. SDI alone on the other hand always performs similarly to or worse than ADI. Regarding model completeness, SADI and ASDI show the highest S/N for $\Omega < 0.2$ and CODI for $\Omega \lessapprox 0.1$.
        
        To quantify the residual speckle noise over the whole FOV, we plotted the noise at different angular separations (see Fig. \ref{FigSNmap}). The noise is given as the root mean square of the pixel values within the reference aperture after PCA, averaged over all reference apertures at the same angular separation (see section \ref{sec::sn_calc}). In all three targets, CODI achieves the lowest noise over the entire angular separation range. The noise values of SADI and ASDI are similar for all three targets and over the entire range of tested angular separations. ADI achieves lower noise than SADI and ASDI at small angular separations ($<$ 0.2 arcsec) in Beta~Pictoris, at large angular separations ($>$ 0.2 arcsec) in HR~8799, and over the entire range in 51~Eridani. Even though it has a considerably lower S/N for all three targets, SDI has a noise level comparable to that of 
CODI in 51~Eridani and that of SADI/ASDI in HR~8799.
    
    For the tests on integration time, field rotation, and wavelength range, various subsets with fewer images and/or wavelength frames than the full dataset were used. Therefore, the optimal number of PCs could not be determined from Fig. \ref{FigDetection}. For all our tests, we chose the 
PC number that achieves the highest S/N over the whole range of tested integration times. The corresponding PC numbers can be seen in Table \ref{tab:S/N}.
    
        The S/N for different integration times of Beta~Pictoris~b and 51~Eridani~b can be seen in Fig. \ref{FigTime}. Only S/N values above 0.5 are displayed. The tests showed that for integration times $t >$ 1500 s, CODI, SADI, ASDI, and ADI approximately follow the photon noise curve with no significant deviation measured. In this case, the tests do not show significant differences in S/N for SADI, ASDI, and CODI. 

        The S/N for different field rotation of Beta~Pictoris~b and 51~Eridani~b 
can be seen in Fig. \ref{FigAngle}. The same PC numbers as for the integration time tests were used (see Table \ref{tab:S/N}). Each subset of Beta~Pictoris contained 76 images and each subset of 51~Eridani contained 15 images. In both targets, a strong increase in S/N with increasing field rotation can be seen for $\Delta \alpha < 1 \mathrm{FWHM}$. After that, the 
increase begins to stagnate and after $\Delta \alpha \gtrapprox 1.5 \mathrm{FWHM}$ no significant increase was measured.
        
        The S/N for different wavelength ranges of Beta~Pictoris~b can be seen in Fig. \ref{FigWave}. The same PC numbers as for the integration time tests were used (see Table \ref{tab:S/N}).
        
        Because the datasets were split into smaller subsets, none of the three dependence tests (integration time, field rotation, and wavelength range) for HR~8799~e achieved S/N of over 5$\sigma$. Therefore, no clear relation could be determined and the dependence analysis of HR~8799~e was not included in this paper.

\section{Discussion}
\label{sec:Discussion}
    \subsection{Differential imaging techniques}
    SADI, ASDI, and CODI all achieve higher S/N compared to using only ADI 
for the three test targets Beta~Pictoris~b, 51~Eridani~b, and HR~8799~e. SDI alone does not result in S/N over 5$\sigma$ in any of the three targets.
    
    We compare our results with the work of \citet{christiaens_separating_2019} on PDS~70. Of the three targets we tested, we focus on Beta~Pictoris~b because its angular separation is closest to PDS~70~b. \citet{christiaens_separating_2019} subsequently used ADI (PCA on concentric annuli of 2 FWHM) and SDI (full-frame PCA). Additionally, these authors only used frames between which the companion would have a rotational offset of at least 1 FWHM for their PSF library. In their analysis, only PCA--ASDI (similar to 
our ASDI) was able to recover a point-like feature; PCA--SADI (similar to our 
SADI) was not. The authors did not apply ADI and SDI simultaneously. In 
our analysis of Beta~Pictoris~b, SADI achieved a higher S/N compared to ASDI. In contrast to our analysis, \citet{christiaens_separating_2019} used the VLT instrument SINFONI \citep{eisenhauer_sinfoni_2003, bonnet_first_2004} which observes in the near-infrared (1.1 - 2.45 $\mu$m; similar to SPHERE/IFS) but has a higher resolution (approximately 2000 in the $J$-band). Furthermore, PDS 70 b has a smaller angular separation (0.193$\pm$0.0049 arcsec) than Beta~Pictoris~b 
(0.249$\pm$0.001 arcsec). Because the quality of SDI strongly depends on the radial displacement (see Eq. \ref{eq:DeltaR}), the different separations can be a dominant factor in the image quality. Furthermore, SINFONI only achieves Strehl ratios of approximately 30\% (whereas SPHERE can go to above 80\% in the J band). This results in a larger FWHM in the SINFONI 
data with respect to the SPHERE data and therefore to smaller S/N at small separations. The observed angle variation strongly influences ADI. Their observations of PDS 70 b have a field rotation of 99.8$^\circ$ while the observations of Beta Pictoris have 41.1$^\circ$. The comparison indicates that the first processing step (ADI in ASDI; SDI in SADI) has a stronger influence on the final images, but this may be limited to this specific case and may not be a general result.
    
    For 51 Eridani b and HR~8799~e, no clear difference between the differential imaging techniques SADI, ASDI, and CODI could be found. \citet{samland_spectral_2017} analysed 51~Eridani~b  using classical SDI followed by ADI (applied using various algorithms). These authors selected reference wavelength channels to improve their SDI performance and achieved a S/N of close to 20, while our SADI achieves a S/N of roughly 10. The improvement could be due to the longer integration time used ($\approx 16384$s) compared to our dataset ($\approx4928$s). Assuming photon-noise-limit behaviour, our dataset would achieve an S/N of roughly 18 and thus be comparable to the results of \citet{samland_spectral_2017} without requiring target-specific adjustments. On the other hand, the difference could also be 
influenced by their in-depth analysis of the different wavelength channels and subsequent improvement of their SDI step.
        
    For all three test objects, CODI achieves slightly higher S/N than SADI or ASDI and requires the lowest model completeness. This can be explained by the higher resolution of the model completeness. The number of PCs used in SADI and ASDI can only be optimised up to $\Delta\Omega = 1/\mathrm{PC}_{max} = 0.026$. The resolution of the model completeness is limited by the maximum number of PCs (see Table \ref{tab:PCmax}). The PC of CODI can be adjusted up to $\Delta\Omega = {1}/(39 \times \mathrm{N}_\mathrm{tot, 
\lambda})$ and in the case of 51 Eridani, this leads to $\Delta\Omega = 
10^{-4}$.
    
    Similar to our CODI, \citet{flasseur_paco_2020} developed PACO--ASDI which exploits the displacement caused by field rotation and wavelength-dependent scaling. In contrast to our work, PACO--ASDI does not use PCA but the PACO algorithm \citep{flasseur_exoplanet_2018}. These latter authors found that using 
angular and radial displacements simultaneously leads to a significant increase in S/N. This indicates that our results could be generalised to other ADI/SDI-based techniques. \citet{flasseur_paco_2020} did not investigate other combinations of ADI and SDI (like SADI and ASDI) and did not test the dependence of  their technique on integration time and wavelength range. 
They did test the dependence on separation, which can be related to field 
rotation. Even though field rotation and angular separation both influence self-subtraction linearly (see Eq. \ref{eq:DeltaRot}), extending the angular separation decreases the influence of the stellar PSF which can  additionally increase the S/N.
    
    Our results were obtained using PCA, but a similar investigation could be done for other techniques like PACO, LOCI, or machine-learning algorithms. All of these algorithms rely on the angular displacement of a planetary companion within the quasi-static PSF halo of the primary star throughout the observing sequence. An additional radial displacement, as used in SDI-based processing methods, can also be used by all the aforementioned algorithms to enhance the diversity of planetary signal and stellar 
residuals. Therefore, an investigation into the quality of combining ADI and SDI could help determine the best observation strategy. The same holds true for our analysis on integration time, field rotation, and wavelength range.

    \subsection{Noise curves}
    The noise curves in Fig. \ref{FigSNmap} show that CODI more effectively reduces the residual speckle noise than the other differential imaging 
techniques. At separations larger than 0.5 arcsec, a steep drop in noise is observed. CODI is therefore expected to perform well at angular separations larger than 0.5 arcsec because self-subtraction effects also weaken with increased angular separation due to larger azimuthal and radial displacements (see Eq. \ref{eq:DeltaR} and \ref{eq:DeltaRot}).
    
    Comparing the noise curves for ADI to SADI/ASDI of the three test targets shows no general trend. ADI can achieve noise values orders of magnitude lower than SADI/ASDI and vise versa. Nevertheless, SADI and ASDI achieve higher S/N in all three targets which can be explained by self-subtraction effects. Ideally, the signal from the  companion is only present in the very last PCs. However, some of its signal will appear in lower order PCs if 
the companion has insufficient displacements between the images. Therefore, if more PCs are used (which leads to lower noise), more of the signal from the  companion is reduced. Selecting the optimal PC number means balancing 
these two effects against each other. SDI never achieves high S/N even though the noise level can be comparable with the other techniques, indicating that SDI cannot efficiently reduce noise before self-subtraction effects start to become dominant.
    
    In most cases, the noise becomes smaller with increasing angular separation, which is expected because the stellar flux decreases. For ADI and CODI in the dataset of Beta~Pictoris, and for SADI and ASDI in the dataset of 51~Eridani, the noise increases for larger angular separations. This increase can be explained by wind-driven halos, which are caused by imperfect corrections of the AO-system due to atmospheric changes attributable to jet streams \citep{cantalloube_origin_2018, cantalloube_wind-driven_2020}. Because these effects are asymmetric, ADI might not correct them well, 
leading to increased noise at larger angular separation.

    \subsection{Dependence tests}
        \subsubsection{Integration time}
        In both  Beta~Pictoris~b and 51~Eridani~b (Fig. \ref{FigTime}), the S/N increase rapidly before following the photon noise curve. We find no significant difference between SADI, ASDI, and CODI. Compared to ADI, all three differential imaging techniques combining ADI and SDI (SADI, ASDI and CODI) require less integration time than ADI to achieve higher S/N.
        
        Similar to our test, \citet{janson_naco-sdi_2007} analysed the integration-time dependence of $\epsilon$ Eridani b using a combination of ADI and 
SDI similar to ASDI. In agreement with our results, these authors found no significant deviation from the photon noise curve with increasing integration time. This dependence is important because some processing algorithms only use a subset of the whole dataset to model the PSF to reduce self-subtraction effects \citep[e.g.][]{zurlo_first_2016, samland_spectral_2017}. Our 
results indicate that the strategic selection of frames can be used as long as the total integration time of the selected frames still exceeds 10$^3$s.

        \subsubsection{Field rotation}
        The results for Beta~Pictoris~b and 51~Eridani~b (Fig. \ref{FigAngle}) show an increased S/N for larger angles if the field rotation is smaller than 1 FWHM. Beyond that, the S/N increase for larger rotations is significantly smaller. This suggests that observations should have a field rotation of at least 1 FWHM at the angular separation of the potential companion. Similar to the integration time analysis, no significant difference between SADI, ASDI, and CODI could be found. Because each angle set can contain different images, the S/N varies slightly depending on the overall quality of the images in the set. This is to be expected and is not caused by the field rotation.
        
        \citet{meshkat_optimized_2013} analysed the S/N dependence of ADI on angular separation. As shown in Eq. \ref{eq:DeltaRot}, the rotational offset depends linearly on both field rotation and angular separation. In accordance with our results, their analysis showed a strong increase in S/N for larger angular separations. For larger angular separations (which correspond to larger angular displacements), these latter authors do not find a significant stagnation of the S/N. Even though field rotation and angular separation both influence self-subtraction linearly (see Eq. \ref{eq:DeltaRot}), varying the angular separation additionally leads to a significant change in the influence of the stellar PSF.

        \subsubsection{Wavelength range}
        For 51~Eridani~b (Fig. \ref{FigWave}), ADI is approximately constant while SADI, ASDI, and CODI improve with increasing wavelength range. This confirms that an increase in observed wavelength range can lead to higher S/N if SDI is used as part of differential imaging techniques.
        
        For Beta~Pictoris~b (Fig. \ref{FigWave}), the S/N of ADI increases with larger wavelength ranges even though the technique itself does not depend 
on the wavelength range. This increase in S/N is explained with the wavelength-dependent brightness of Beta~Pictoris~b. We compared the S/N increase of ADI with the spectra of Beta~Pictoris~b obtained by \citet{chilcote_124upmum_2017}. These latter authors measured an approximately linear increase for the wavelength range 0.95 - 1.35 $\mu$m, which explains the S/N increase that we measure for 
ADI. SADI and ASDI only achieved reasonable detections (S/N$>5$) if the whole wavelength range was used, and their performance is worse than ADI. Analysing the processed images shows strong self-subtraction features at the position of Beta~Pictoris~b which explains the reduced S/N. Images processed with CODI did not show signs of significant self-subtraction and achieve a reasonably high S/N in this test. This result suggests that CODI can be more robust in cases where self-subtraction is dominant. A further 
investigation of the effects of injecting a fake companion could be used to study this phenomenon in more detail.

\section{Conclusion}
\label{sec:summary}
    We tested the performance of ADI, SDI, SADI (using SDI, then ADI), ASDI (using ADI, then SDI), and CODI (using ADI and SDI simultaneously) on three different targets: Beta~Pictoris~b, 51~Eridani~b, and HR~8799~e. Beta~Pictoris~b and 51 Eridani b were used to test the dependence of the techniques on integration time, field rotation, and observed wavelength range. The differential imaging techniques used in this paper are available within the PynPoint library.
    
    For these three targets, using a combination of ADI and SDI consistently achieved better results than using ADI alone. SADI and ASDI performed 
similarly but target-specific differences can occur. In all tests, CODI achieved the highest S/N and constantly performed slightly better than SADI or ASDI.
    
    We analysed the dependence of differential imaging techniques on integration time by splitting the datasets of our test targets into subsets of different total integration time. In this test, SADI, ASDI, and CODI all showed a photon-noise-limited behaviour for integration times of $>$ 1500 s if all images were considered. Choosing frames selectively to reduce self-subtraction might decrease this time limit.
    
    We analysed the dependence of differential imaging techniques on field rotation by splitting the datasets of our test targets into subsets with different observed rotation but with the same integration time. In this 
test, SADI, ASDI, and CODI all showed a strong increase of S/N up to displacements of 1 FWHM. After that, increasing the field rotation only led to a small increase in S/N.

    We analysed the dependence of differential imaging techniques on the observed wavelength range by splitting the datasets of our test targets into subsets with different wavelength ranges. Our tests show a general increase in S/N with increasing wavelength range, but no clear relation could be found. Furthermore, the spectra of a given exoplanet influence the results because its brightness can vary substantially in different wavelength channels.
    
    To further investigate these techniques, SADI and ASDI should be compared to CODI while using different numbers of PCs in SDI and ADI. In combination with analysing more known exoplanets and brown dwarfs, this would help to decipher whether or not our findings can be generalised to a wider selection of targets. Unfortunately, both tasks are immensely computationally intensive. Computationally optimising the techniques could 
help to reduce the intensity and would make a large-scale search for exoplanets more feasible. Furthermore, the extraction of spectra using CODI should be investigated further. This requires careful treatment of the different wavelengths because of cross talk between adjacent spectral channels \citep{greco_measurement_2016}.

        Our analysis of the differential imaging techniques ADI, SDI, CODI, SADI, 
and ASDI shows that using SDI in combination with ADI leads to higher S/N for exoplanet detections than using only ADI or SDI. Overall, CODI achieved the highest S/N in all three targets. This can be explained with the higher model completeness resolution which enables this technique to more precisely model the stellar PSF. We suggest that future observations should first be evaluated with CODI if it is not feasible to apply all of them. The observed wavelength range should be maximised (e.g. for SPHERE/IFS use 0.95 - 1.65 $\mu$m) and the field rotation should cover at least 2 FWHM at the angular separation 
where exoplanets are expected. An integration time of 10$^3$s is sufficient to reach photon noise behaviour.

\begin{acknowledgements}
To achieve the scientific results presented in this article we made use of the Python, especially the SciPy \citep{virtanen_scipy_2020}, NumPy \citep{harris_array_2020}, Matplotlib \citep{hunter_matplotlib_2007}, emcee \citep{foreman-mackey_emcee_2013}, scikit-image \citep{walt_scikit-image_2014}, scikit-learn \citep{pedregosa_scikit-learn_2018}, photutils \citep{bradley_photutils_2016}, and astropy \citep{astropy_collaboration_astropy_2013, astropy_collaboration_astropy_2018} packages.

Part of this work has been carried out within the framework of the National Centre of Competence in Research PlanetS supported by the Swiss National Science Foundation. SPQ acknowledges the financial support of the SNSF.
\end{acknowledgements}

\bibliographystyle{aa} 
\bibliography{bibliography} 

\begin{thebibliography}{63}
\expandafter\ifx\csname natexlab\endcsname\relax\def\natexlab#1{#1}\fi

\bibitem[{Amara \& Quanz(2012)}]{amara_pynpoint_2012}
Amara, A. \& Quanz, S.~P. 2012, Monthly Notices of the Royal Astronomical
  Society, 427, 948

\bibitem[{Apai {et~al.}(2016)Apai, Kasper, Skemer, Hanson, Lagrange, Biller,
  Bonnefoy, Buenzli, \& Vigan}]{apai_high-cadence_2016}
Apai, D., Kasper, M., Skemer, A., {et~al.} 2016, The Astrophysical Journal,
  820, 40

\bibitem[{{Astropy Collaboration} {et~al.}(2018){Astropy Collaboration},
  Price-Whelan, Sipőcz, Günther, Lim, Crawford, Conseil, Shupe, Craig,
  Dencheva, Ginsburg, VanderPlas, Bradley, Pérez-Suárez, de~Val-Borro,
  Aldcroft, Cruz, Robitaille, Tollerud, Ardelean, Babej, Bach, Bachetti,
  Bakanov, Bamford, Barentsen, Barmby, Baumbach, Berry, Biscani, Boquien,
  Bostroem, Bouma, Brammer, Bray, Breytenbach, Buddelmeijer, Burke, Calderone,
  Cano~Rodríguez, Cara, Cardoso, Cheedella, Copin, Corrales, Crichton,
  D'Avella, Deil, Depagne, Dietrich, Donath, Droettboom, Earl, Erben, Fabbro,
  Ferreira, Finethy, Fox, Garrison, Gibbons, Goldstein, Gommers, Greco,
  Greenfield, Groener, Grollier, Hagen, Hirst, Homeier, Horton, Hosseinzadeh,
  Hu, Hunkeler, Ivezić, Jain, Jenness, Kanarek, Kendrew, Kern, Kerzendorf,
  Khvalko, King, Kirkby, Kulkarni, Kumar, Lee, Lenz, Littlefair, Ma, Macleod,
  Mastropietro, McCully, Montagnac, Morris, Mueller, Mumford, Muna, Murphy,
  Nelson, Nguyen, Ninan, Nöthe, Ogaz, Oh, Parejko, Parley, Pascual, Patil,
  Patil, Plunkett, Prochaska, Rastogi, Reddy~Janga, Sabater, Sakurikar,
  Seifert, Sherbert, Sherwood-Taylor, Shih, Sick, Silbiger, Singanamalla,
  Singer, Sladen, Sooley, Sornarajah, Streicher, Teuben, Thomas, Tremblay,
  Turner, Terrón, van Kerkwijk, de~la Vega, Watkins, Weaver, Whitmore,
  Woillez, Zabalza, \& {Astropy
  Contributors}}]{astropy_collaboration_astropy_2018}
{Astropy Collaboration}, Price-Whelan, A.~M., Sipőcz, B.~M., {et~al.} 2018,
  The Astronomical Journal, 156, 123

\bibitem[{{Astropy Collaboration} {et~al.}(2013){Astropy Collaboration},
  Robitaille, Tollerud, Greenfield, Droettboom, Bray, Aldcroft, Davis,
  Ginsburg, Price-Whelan, Kerzendorf, Conley, Crighton, Barbary, Muna,
  Ferguson, Grollier, Parikh, Nair, Unther, Deil, Woillez, Conseil, Kramer,
  Turner, Singer, Fox, Weaver, Zabalza, Edwards, Azalee~Bostroem, Burke, Casey,
  Crawford, Dencheva, Ely, Jenness, Labrie, Lim, Pierfederici, Pontzen, Ptak,
  Refsdal, Servillat, \& Streicher}]{astropy_collaboration_astropy_2013}
{Astropy Collaboration}, Robitaille, T.~P., Tollerud, E.~J., {et~al.} 2013,
  Astronomy and Astrophysics, 558, A33

\bibitem[{Beuzit {et~al.}(2019)Beuzit, Vigan, Mouillet, Dohlen, Gratton,
  Boccaletti, Sauvage, Schmid, Langlois, Petit, Baruffolo, Feldt, Milli,
  Wahhaj, Abe, Anselmi, Antichi, Barette, Baudrand, Baudoz, Bazzon, Bernardi,
  Blanchard, Brast, Bruno, Buey, Carbillet, Carle, Cascone, Chapron, Charton,
  Chauvin, Claudi, Costille, Caprio, Boer, Delboulbé, Desidera, Dominik,
  Downing, Dupuis, Fabron, Fantinel, Farisato, Feautrier, Fedrigo, Fusco,
  Gigan, Ginski, Girard, Giro, Gisler, Gluck, Gry, Henning, Hubin, Hugot,
  Incorvaia, Jaquet, Kasper, Lagadec, Lagrange, Coroller, Mignant, Ruyet,
  Lessio, Lizon, Llored, Lundin, Madec, Magnard, Marteaud, Martinez, Maurel,
  Ménard, Mesa, Möller-Nilsson, Moulin, Moutou, Origné, Parisot, Pavlov,
  Perret, Pragt, Puget, Rabou, Ramos, Reess, Rigal, Rochat, Roelfsema, Rousset,
  Roux, Saisse, Salasnich, Santambrogio, Scuderi, Segransan, Sevin,
  Siebenmorgen, Soenke, Stadler, Suarez, Tiphène, Turatto, Udry, Vakili,
  Waters, Weber, Wildi, Zins, \& Zurlo}]{beuzit_sphere_2019}
Beuzit, J.-L., Vigan, A., Mouillet, D., {et~al.} 2019, Astronomy \&
  Astrophysics, 631, A155, publisher: EDP Sciences

\bibitem[{Biller {et~al.}(2013)Biller, Liu, Wahhaj, Nielsen, Hayward, Males,
  Skemer, Close, Chun, Ftaclas, Clarke, Thatte, Shkolnik, Reid, Hartung, Boss,
  Lin, Alencar, Pino, Gregorio-Hetem, \& Toomey}]{biller_gemininici_2013}
Biller, B.~A., Liu, M.~C., Wahhaj, Z., {et~al.} 2013, The Astrophysical
  Journal, 777, 160, publisher: IOP Publishing

\bibitem[{Bonnefoy {et~al.}(2013)Bonnefoy, Boccaletti, Lagrange, Allard,
  Mordasini, Beust, Chauvin, Girard, Homeier, Apai, Lacour, \&
  Rouan}]{bonnefoy_near-infrared_2013}
Bonnefoy, M., Boccaletti, A., Lagrange, A.-M., {et~al.} 2013, Astronomy \&
  Astrophysics, 555, A107

\bibitem[{Bonnet {et~al.}(2004)Bonnet, Abuter, Baker, Bornemann, Brown,
  Castillo, Conzelmann, Damster, Davies, Delabre, Donaldson, Dumas, Eisenhauer,
  Elswijk, Fedrigo, Finger, Gemperlein, Genzel, Gilbert, Gillet, Goldbrunner,
  Horrobin, Ter~Horst, Huber, Hubin, Iserlohe, Kaufer, Kissler-Patig, Kragt,
  Kroes, Lehnert, Lieb, Liske, Lizon, Lutz, Modigliani, Monnet, Nesvadba,
  Patig, Pragt, Reunanen, Röhrle, Rossi, Schmutzer, Schoenmaker, Schreiber,
  Stroebele, Szeifert, Tacconi, Tecza, Thatte, Tordo, van~der Werf, \&
  Weisz}]{bonnet_first_2004}
Bonnet, H., Abuter, R., Baker, A., {et~al.} 2004, The Messenger, 117, 17

\bibitem[{Bradley {et~al.}(2016)Bradley, Sipocz, Robitaille, Tollerud, Deil,
  Vinícius, Barbary, Günther, Bostroem, Droettboom, Bray, Bratholm,
  Pickering, Craig, Pascual, Greco, Donath, Kerzendorf, Littlefair, Barentsen,
  D'Eugenio, \& Weaver}]{bradley_photutils_2016}
Bradley, L., Sipocz, B., Robitaille, T., {et~al.} 2016, Astrophysics Source
  Code Library, ascl:1609.011

\bibitem[{Cantalloube {et~al.}(2020)Cantalloube, Farley, Milli, Bharmal,
  Brandner, Correia, Dohlen, Henning, Osborn, Por, Suárez~Valles, \&
  Vigan}]{cantalloube_wind-driven_2020}
Cantalloube, F., Farley, O. J.~D., Milli, J., {et~al.} 2020, Astronomy and
  Astrophysics, 638, A98

\bibitem[{Cantalloube {et~al.}(2015)Cantalloube, Mouillet, Mugnier, Milli,
  Absil, Gomez~Gonzalez, Chauvin, Beuzit, \& Cornia}]{cantalloube_direct_2015}
Cantalloube, F., Mouillet, D., Mugnier, L.~M., {et~al.} 2015, Astronomy \&
  Astrophysics, 582, A89

\bibitem[{Cantalloube {et~al.}(2018)Cantalloube, Por, Dohlen, Sauvage, Vigan,
  Kasper, Bharmal, Henning, Brandner, Milli, Correia, \&
  Fusco}]{cantalloube_origin_2018}
Cantalloube, F., Por, E.~H., Dohlen, K., {et~al.} 2018, Astronomy and
  Astrophysics, 620, L10

\bibitem[{Chilcote {et~al.}(2017)Chilcote, Pueyo, Rosa, Vargas, Macintosh,
  Bailey, Barman, Bauman, Bruzzone, Bulger, Burrows, Cardwell, Chen, Cotten,
  Dillon, Doyon, Draper, Duchêne, Dunn, Erikson, Fitzgerald, Follette, Gavel,
  Goodsell, Graham, Greenbaum, Hartung, Hibon, Hung, Ingraham, Kalas,
  Konopacky, Larkin, Maire, Marchis, Marley, Marois, Metchev, Millar-Blanchaer,
  Morzinski, Nielsen, Norton, Oppenheimer, Palmer, Patience, Perrin, Poyneer,
  Rajan, Rameau, Rantakyrö, Sadakuni, Saddlemyer, Savransky, Schneider, Serio,
  Sivaramakrishnan, Song, Soummer, Thomas, Wallace, Wang, Ward-Duong,
  Wiktorowicz, \& Wolff}]{chilcote_124upmum_2017}
Chilcote, J., Pueyo, L., Rosa, R. J.~D., {et~al.} 2017, The Astronomical
  Journal, 153, 182, publisher: American Astronomical Society

\bibitem[{Christiaens {et~al.}(2019)Christiaens, Casassus, Absil, Cantalloube,
  Gonzalez, Girard, Ramirez, Pairet, Salinas, Price, Pinte, Quanz, Jordan,
  Mawet, \& Wahhaj}]{christiaens_separating_2019}
Christiaens, V., Casassus, S., Absil, O., {et~al.} 2019, Monthly Notices of the
  Royal Astronomical Society, 486, 5819, arXiv: 1905.01860

\bibitem[{Claudi {et~al.}(2008)Claudi, Turatto, Gratton, Antichi, Bonavita,
  Bruno, Cascone, Caprio, Desidera, Giro, Mesa, Scuderi, Dohlen, Beuzit, \&
  Puget}]{claudi_sphere_2008}
Claudi, R.~U., Turatto, M., Gratton, R.~G., {et~al.} 2008, in Ground-based and
  {Airborne} {Instrumentation} for {Astronomy} {II}, Vol. 7014 (International
  Society for Optics and Photonics), 70143E

\bibitem[{Eisenhauer {et~al.}(2003)Eisenhauer, Abuter, Bickert,
  Biancat-Marchet, Bonnet, Brynnel, Conzelmann, Delabre, Donaldson, Farinato,
  Fedrigo, Genzel, Hubin, Iserlohe, Kasper, Kissler-Patig, Monnet, Roehrle,
  Schreiber, Stroebele, Tecza, Thatte, \& Weisz}]{eisenhauer_sinfoni_2003}
Eisenhauer, F., Abuter, R., Bickert, K., {et~al.} 2003, The international
  society for optics and photonics, 4841, 1548, conference Name: Instrument
  Design and Performance for Optical/Infrared Ground-based Telescopes Place:
  eprint: arXiv:astro-ph/0306191

\bibitem[{Flasseur {et~al.}(2018)Flasseur, Denis, Thiébaut, \&
  Langlois}]{flasseur_exoplanet_2018}
Flasseur, O., Denis, L., Thiébaut, E., \& Langlois, M. 2018, Astronomy and
  Astrophysics, 618, A138

\bibitem[{Flasseur {et~al.}(2020)Flasseur, Denis, Thiébaut, \&
  Langlois}]{flasseur_paco_2020}
Flasseur, O., Denis, L., Thiébaut, E., \& Langlois, M. 2020, Astronomy and
  Astrophysics, 637, A9

\bibitem[{Foreman-Mackey {et~al.}(2013)Foreman-Mackey, Hogg, Lang, \&
  Goodman}]{foreman-mackey_emcee_2013}
Foreman-Mackey, D., Hogg, D.~W., Lang, D., \& Goodman, J. 2013, Publications of
  the Astronomical Society of the Pacific, 125, 306

\bibitem[{Freudling {et~al.}(2013)Freudling, Romaniello, Bramich, Ballester,
  Forchi, Garcia-Dablo, Moehler, \& Neeser}]{freudling_automated_2013}
Freudling, W., Romaniello, M., Bramich, D.~M., {et~al.} 2013, Astronomy \&
  Astrophysics, 559, A96, arXiv: 1311.5411

\bibitem[{Galicher {et~al.}(2018)Galicher, Boccaletti, Mesa, Delorme, Gratton,
  Langlois, Lagrange, Maire, Coroller, Chauvin, Biller, Cantalloube, Janson,
  Lagadec, Meunier, Vigan, Hagelberg, Bonnefoy, Zurlo, Rocha, Maurel, Jaquet,
  Buey, \& Weber}]{galicher_astrometric_2018}
Galicher, R., Boccaletti, A., Mesa, D., {et~al.} 2018, Astronomy \&
  Astrophysics, 615, A92, publisher: EDP Sciences

\bibitem[{Gebhard {et~al.}(2020)Gebhard, Bonse, Quanz, \&
  Schölkopf}]{gebhard_physically_2020}
Gebhard, T.~D., Bonse, M.~J., Quanz, S.~P., \& Schölkopf, B. 2020, arXiv
  e-prints, 2010, arXiv:2010.05591

\bibitem[{Gonzalez {et~al.}(2018)Gonzalez, Absil, \& van
  Droogenbroeck}]{gonzalez_supervised_2018}
Gonzalez, C. A.~G., Absil, O., \& van Droogenbroeck, M. 2018, Astronomy \&
  Astrophysics, 613, A71, arXiv: 1712.02841

\bibitem[{Gonzalez {et~al.}(2017)Gonzalez, Wertz, Absil, Christiaens, Defrère,
  Mawet, Milli, Absil, Droogenbroeck, Cantalloube, Hinz, Skemer, Karlsson, \&
  Surdej}]{gonzalez_vip_2017}
Gonzalez, C. A.~G., Wertz, O., Absil, O., {et~al.} 2017, The Astronomical
  Journal, 154, 7, publisher: American Astronomical Society

\bibitem[{Greco \& Brandt(2016)}]{greco_measurement_2016}
Greco, J.~P. \& Brandt, T.~D. 2016, The Astrophysical Journal, 833, 134

\bibitem[{Haffert {et~al.}(2019)Haffert, Bohn, de~Boer, Snellen, Brinchmann,
  Girard, Keller, \& Bacon}]{haffert_two_2019}
Haffert, S.~Y., Bohn, A.~J., de~Boer, J., {et~al.} 2019, Nature Astronomy, 3,
  749

\bibitem[{Harris {et~al.}(2020)Harris, Millman, van~der Walt, Gommers,
  Virtanen, Cournapeau, Wieser, Taylor, Berg, Smith, Kern, Picus, Hoyer, van
  Kerkwijk, Brett, Haldane, del Río, Wiebe, Peterson, Gérard-Marchant,
  Sheppard, Reddy, Weckesser, Abbasi, Gohlke, \& Oliphant}]{harris_array_2020}
Harris, C.~R., Millman, K.~J., van~der Walt, S.~J., {et~al.} 2020, Nature, 585,
  357, number: 7825 Publisher: Nature Publishing Group

\bibitem[{Heinze {et~al.}(2010)Heinze, Hinz, Kenworthy, Meyer, Sivanandam, \&
  Miller}]{heinze_constraints_2010}
Heinze, A.~N., Hinz, P.~M., Kenworthy, M., {et~al.} 2010, The Astrophysical
  Journal, 714, 1570

\bibitem[{Hunter(2007)}]{hunter_matplotlib_2007}
Hunter, J.~D. 2007, Computing in Science and Engineering, 9, 90

\bibitem[{Janson {et~al.}(2007)Janson, Brandner, Henning, Lenzen, McArthur,
  Benedict, Reffert, Nielsen, Close, Biller, Kellner, Günther, Hatzes,
  Masciadri, Geissler, \& Hartung}]{janson_naco-sdi_2007}
Janson, M., Brandner, W., Henning, T., {et~al.} 2007, The Astronomical Journal,
  133, 2442, publisher: IOP Publishing

\bibitem[{Jee {et~al.}(2007)Jee, Blakeslee, Sirianni, Martel, White, \&
  Ford}]{jee_principal_2007}
Jee, M.~J., Blakeslee, J.~P., Sirianni, M., {et~al.} 2007, Publications of the
  Astronomical Society of the Pacific, 119, 1403

\bibitem[{Keppler {et~al.}(2018)Keppler, Benisty, Müller, Henning, van Boekel,
  Cantalloube, Ginski, van Holstein, Maire, Pohl, Samland, Avenhaus, Baudino,
  Boccaletti, de~Boer, Bonnefoy, Chauvin, Desidera, Langlois, Lazzoni, Marleau,
  Mordasini, Pawellek, Stolker, Vigan, Zurlo, Birnstiel, Brandner, Feldt,
  Flock, Girard, Gratton, Hagelberg, Isella, Janson, Juhasz, Kemmer, Kral,
  Lagrange, Launhardt, Matter, Ménard, Milli, Mollière, Olofsson, Pérez,
  Pinilla, Pinte, Quanz, Schmidt, Udry, Wahhaj, Williams, Buenzli, Cudel,
  Dominik, Galicher, Kasper, Lannier, Mesa, Mouillet, Peretti, Perrot, Salter,
  Sissa, Wildi, Abe, Antichi, Augereau, Baruffolo, Baudoz, Bazzon, Beuzit,
  Blanchard, Brems, Buey, De~Caprio, Carbillet, Carle, Cascone, Cheetham,
  Claudi, Costille, Delboulbé, Dohlen, Fantinel, Feautrier, Fusco, Giro,
  Gluck, Gry, Hubin, Hugot, Jaquet, Le~Mignant, Llored, Madec, Magnard,
  Martinez, Maurel, Meyer, Möller-Nilsson, Moulin, Mugnier, Origné, Pavlov,
  Perret, Petit, Pragt, Puget, Rabou, Ramos, Rigal, Rochat, Roelfsema, Rousset,
  Roux, Salasnich, Sauvage, Sevin, Soenke, Stadler, Suarez, Turatto, \&
  Weber}]{keppler_discovery_2018}
Keppler, M., Benisty, M., Müller, A., {et~al.} 2018, Astronomy and
  Astrophysics, 617, A44

\bibitem[{Lafreniere {et~al.}(2007)Lafreniere, Marois, Doyon, Nadeau, \&
  Artigau}]{lafreniere_new_2007}
Lafreniere, D., Marois, C., Doyon, R., Nadeau, D., \& Artigau, E. 2007, The
  Astrophysical Journal, 660, 770, arXiv: astro-ph/0702697

\bibitem[{Lagrange {et~al.}(2019)Lagrange, Boccaletti, Langlois, Chauvin,
  Gratton, Beust, Desidera, Milli, Bonnefoy, Cheetham, Feldt, Meyer, Vigan,
  Biller, Bonavita, Baudino, Cantalloube, Cudel, Daemgen, Delorme, D'Orazi,
  Girard, Fontanive, Hagelberg, Janson, Keppler, Koypitova, Galicher, Lannier,
  Le~Coroller, Ligi, Maire, Mesa, Messina, Müeller, Peretti, Perrot, Rouan,
  Salter, Samland, Schmidt, Sissa, Zurlo, Beuzit, Mouillet, Dominik, Henning,
  Lagadec, Ménard, Schmid, Turatto, Udry, Bohn, Charnay, Gomez~Gonzales, Gry,
  Kenworthy, Kral, Mordasini, Moutou, van~der Plas, Schlieder, Abe, Antichi,
  Baruffolo, Baudoz, Baudrand, Blanchard, Bazzon, Buey, Carbillet, Carle,
  Charton, Cascone, Claudi, Costille, Deboulbe, De~Caprio, Dohlen, Fantinel,
  Feautrier, Fusco, Gigan, Giro, Gisler, Gluck, Hubin, Hugot, Jaquet, Kasper,
  Madec, Magnard, Martinez, Maurel, Le~Mignant, Möller-Nilsson, Llored,
  Moulin, Origné, Pavlov, Perret, Petit, Pragt, Szulagyi, \&
  Wildi}]{lagrange_post-conjunction_2019}
Lagrange, A.-M., Boccaletti, A., Langlois, M., {et~al.} 2019, Astronomy and
  Astrophysics, 621, L8

\bibitem[{Lagrange {et~al.}(2010)Lagrange, Bonnefoy, Chauvin, Apai, Ehrenreich,
  Boccaletti, Gratadour, Rouan, Mouillet, Lacour, \&
  Kasper}]{lagrange_giant_2010}
Lagrange, A.-M., Bonnefoy, M., Chauvin, G., {et~al.} 2010, Science, 329, 57,
  publisher: American Association for the Advancement of Science Section:
  Report

\bibitem[{Lenzen {et~al.}(2003)Lenzen, Hartung, Brandner, Finger, Hubin,
  Lacombe, Lagrange, Lehnert, Moorwood, \& Mouillet}]{lenzen_naos-conica_2003}
Lenzen, R., Hartung, M., Brandner, W., {et~al.} 2003, in Instrument {Design}
  and {Performance} for {Optical}/{Infrared} {Ground}-based {Telescopes}, Vol.
  4841 (International Society for Optics and Photonics), 944--952

\bibitem[{Macintosh {et~al.}(2015)Macintosh, Graham, Barman, Rosa, Konopacky,
  Marley, Marois, Nielsen, Pueyo, Rajan, Rameau, Saumon, Wang, Patience,
  Ammons, Arriaga, Artigau, Beckwith, Brewster, Bruzzone, Bulger, Burningham,
  Burrows, Chen, Chiang, Chilcote, Dawson, Dong, Doyon, Draper, Duchêne,
  Esposito, Fabrycky, Fitzgerald, Follette, Fortney, Gerard, Goodsell,
  Greenbaum, Hibon, Hinkley, Cotten, Hung, Ingraham, Johnson-Groh, Kalas,
  Lafreniere, Larkin, Lee, Line, Long, Maire, Marchis, Matthews, Max, Metchev,
  Millar-Blanchaer, Mittal, Morley, Morzinski, Murray-Clay, Oppenheimer,
  Palmer, Patel, Perrin, Poyneer, Rafikov, Rantakyrö, Rice, Rojo, Rudy,
  Ruffio, Ruiz, Sadakuni, Saddlemyer, Salama, Savransky, Schneider,
  Sivaramakrishnan, Song, Soummer, Thomas, Vasisht, Wallace, Ward-Duong,
  Wiktorowicz, Wolff, \& Zuckerman}]{macintosh_discovery_2015}
Macintosh, B., Graham, J.~R., Barman, T., {et~al.} 2015, Science, 350, 64,
  publisher: American Association for the Advancement of Science Section:
  Report

\bibitem[{Marois {et~al.}(2014)Marois, Correia, Galicher, Ingraham, Macintosh,
  Currie, \& Rosa}]{marois_gpi_2014}
Marois, C., Correia, C., Galicher, R., {et~al.} 2014, in Adaptive {Optics}
  {Systems} {IV}, Vol. 9148 (International Society for Optics and Photonics),
  91480U

\bibitem[{Marois {et~al.}(2013)Marois, Correia, Véran, \&
  Currie}]{marois_tloci_2013}
Marois, C., Correia, C., Véran, J.-P., \& Currie, T. 2013, Proceedings of the
  International Astronomical Union, 8, 48, publisher: Cambridge University
  Press

\bibitem[{Marois {et~al.}(2006)Marois, Lafreniere, Doyon, Macintosh, \&
  Nadeau}]{marois_angular_2006}
Marois, C., Lafreniere, D., Doyon, R., Macintosh, B., \& Nadeau, D. 2006, The
  Astrophysical Journal, 641, 556

\bibitem[{Marois {et~al.}(2010)Marois, Zuckerman, Konopacky, Macintosh, \&
  Barman}]{marois_images_2010}
Marois, C., Zuckerman, B., Konopacky, Q.~M., Macintosh, B., \& Barman, T. 2010,
  Nature, 468, 1080, number: 7327 Publisher: Nature Publishing Group

\bibitem[{Mawet {et~al.}(2014)Mawet, Milli, Wahhaj, Pelat, Absil, Delacroix,
  Boccaletti, Kasper, Kenworthy, Marois, Mennesson, \&
  Pueyo}]{mawet_fundamental_2014}
Mawet, D., Milli, J., Wahhaj, Z., {et~al.} 2014, The Astrophysical Journal,
  792, 97

\bibitem[{Meshkat {et~al.}(2013)Meshkat, Kenworthy, Quanz, \&
  Amara}]{meshkat_optimized_2013}
Meshkat, T., Kenworthy, M., Quanz, S.~P., \& Amara, A. 2013, The Astrophysical
  Journal, 780, 17, arXiv: 1310.8577

\bibitem[{Mugnier {et~al.}(2009)Mugnier, Cornia, Sauvage, Rousset, Fusco, \&
  Védrenne}]{mugnier_optimal_2009}
Mugnier, L.~M., Cornia, A., Sauvage, J.-F., {et~al.} 2009, JOSA A, 26, 1326,
  publisher: Optical Society of America

\bibitem[{Nielsen {et~al.}(2013)Nielsen, Liu, Wahhaj, Biller, Hayward, Close,
  Males, Skemer, Chun, Ftaclas, Alencar, Artymowicz, Boss, Clarke, Pino,
  Gregorio-Hetem, Hartung, Ida, Kuchner, Lin, Reid, Shkolnik, Tecza, Thatte, \&
  Toomey}]{nielsen_gemini_2013}
Nielsen, E.~L., Liu, M.~C., Wahhaj, Z., {et~al.} 2013, The Astrophysical
  Journal, 776, 4, publisher: IOP Publishing

\bibitem[{Nielsen {et~al.}(2019)Nielsen, Rosa, Macintosh, Wang, Ruffio, Chiang,
  Marley, Saumon, Savransky, Ammons, Bailey, Barman, Blain, Bulger, Burrows,
  Chilcote, Cotten, Czekala, Doyon, Duchêne, Esposito, Fabrycky, Fitzgerald,
  Follette, Fortney, Gerard, Goodsell, Graham, Greenbaum, Hibon, Hinkley,
  Hirsch, Hom, Hung, Dawson, Ingraham, Kalas, Konopacky, Larkin, Lee, Lin,
  Maire, Marchis, Marois, Metchev, Millar-Blanchaer, Morzinski, Oppenheimer,
  Palmer, Patience, Perrin, Poyneer, Pueyo, Rafikov, Rajan, Rameau, Rantakyrö,
  Ren, Schneider, Sivaramakrishnan, Song, Soummer, Tallis, Thomas, Ward-Duong,
  \& Wolff}]{nielsen_gemini_2019}
Nielsen, E.~L., Rosa, R. J.~D., Macintosh, B., {et~al.} 2019, The Astronomical
  Journal, 158, 13, publisher: American Astronomical Society

\bibitem[{Pedregosa {et~al.}(2018)Pedregosa, Varoquaux, Gramfort, Michel,
  Thirion, Grisel, Blondel, Müller, Nothman, Louppe, Prettenhofer, Weiss,
  Dubourg, Vanderplas, Passos, Cournapeau, Brucher, Perrot, \&
  Duchesnay}]{pedregosa_scikit-learn_2018}
Pedregosa, F., Varoquaux, G., Gramfort, A., {et~al.} 2018, arXiv:1201.0490
  [cs], arXiv: 1201.0490

\bibitem[{Racine {et~al.}(1999)Racine, Walker, Nadeau, Doyon, \&
  Marois}]{racine_speckle_1999}
Racine, R., Walker, G. A.~H., Nadeau, D., Doyon, R., \& Marois, C. 1999,
  Publications of the Astronomical Society of the Pacific, 111, 587, publisher:
  IOP Publishing

\bibitem[{Rosa {et~al.}(2015)Rosa, Nielsen, Blunt, Graham, Konopacky, Marois,
  Pueyo, Rameau, Ryan, Wang, Bailey, Chontos, Fabrycky, Follette, Macintosh,
  Marchis, Ammons, Arriaga, Chilcote, Cotten, Doyon, Duchêne, Esposito,
  Fitzgerald, Gerard, Goodsell, Greenbaum, Hibon, Ingraham, Johnson-Groh,
  Kalas, Lafrenière, Maire, Metchev, Millar-Blanchaer, Morzinski, Oppenheimer,
  Patel, Patience, Perrin, Rajan, Rantakyrö, Ruffio, Schneider,
  Sivaramakrishnan, Song, Tran, Vasisht, Ward-Duong, \&
  Wolff}]{rosa_astrometric_2015}
Rosa, R. J.~D., Nielsen, E.~L., Blunt, S.~C., {et~al.} 2015, The Astrophysical
  Journal, 814, L3

\bibitem[{Rousset {et~al.}(2003)Rousset, Lacombe, Puget, Hubin, Gendron, Fusco,
  Arsenault, Charton, Feautrier, Gigan, Kern, Lagrange, Madec, Mouillet,
  Rabaud, Rabou, Stadler, \& Zins}]{rousset_naos--first_2003}
Rousset, G., Lacombe, F., Puget, P., {et~al.} 2003, in Adaptive {Optical}
  {System} {Technologies} {II}, Vol. 4839 (International Society for Optics and
  Photonics), 140--149

\bibitem[{Samland {et~al.}(2020)Samland, Bouwman, Hogg, Brandner, Henning, \&
  Janson}]{samland_trap_2020}
Samland, M., Bouwman, J., Hogg, D.~W., {et~al.} 2020, arXiv e-prints, 2011,
  arXiv:2011.12311

\bibitem[{Samland {et~al.}(2017)Samland, Mollière, Bonnefoy, Maire,
  Cantalloube, Cheetham, Mesa, Gratton, Biller, Wahhaj, Bouwman, Brandner,
  Melnick, Carson, Janson, Henning, Homeier, Mordasini, Langlois, Quanz, van
  Boekel, Zurlo, Schlieder, Avenhaus, Beuzit, Boccaletti, Bonavita, Chauvin,
  Claudi, Cudel, Desidera, Feldt, Fusco, Galicher, Kopytova, Lagrange,
  Le~Coroller, Martinez, Moeller-Nilsson, Mouillet, Mugnier, Perrot, Sevin,
  Sissa, Vigan, \& Weber}]{samland_spectral_2017}
Samland, M., Mollière, P., Bonnefoy, M., {et~al.} 2017, Astronomy \&
  Astrophysics, 603, A57

\bibitem[{Soummer {et~al.}(2012)Soummer, Pueyo, \&
  Larkin}]{soummer_detection_2012}
Soummer, R., Pueyo, L., \& Larkin, J. 2012, The Astrophysical Journal Letters,
  755, L28

\bibitem[{Stolker {et~al.}(2019)Stolker, Bonse, Quanz, Amara, Cugno, Bohn, \&
  Boehle}]{stolker_pynpoint_2019}
Stolker, T., Bonse, M.~J., Quanz, S.~P., {et~al.} 2019, Astronomy and
  Astrophysics, 621, A59

\bibitem[{Thatte {et~al.}(2007)Thatte, Abuter, Tecza, Nielsen, Clarke, \&
  Close}]{thatte_very_2007}
Thatte, N., Abuter, R., Tecza, M., {et~al.} 2007, Monthly Notices of the Royal
  Astronomical Society, 378, 1229

\bibitem[{Vigan {et~al.}(2020)Vigan, Fontanive, Meyer, Biller, Bonavita, Feldt,
  Desidera, Marleau, Emsenhuber, Galicher, Rice, Forgan, Mordasini, Gratton,
  Coroller, Maire, Cantalloube, Chauvin, Cheetham, Hagelberg, Lagrange,
  Langlois, Bonnefoy, Beuzit, Boccaletti, D'Orazi, Delorme, Dominik, Henning,
  Janson, Lagadec, Lazzoni, Ligi, Menard, Mesa, Messina, Moutou, Müller,
  Perrot, Samland, Schmid, Schmidt, Sissa, Turatto, Udry, Zurlo, Abe, Antichi,
  Asensio-Torres, Baruffolo, Baudoz, Baudrand, Bazzon, Blanchard, Bohn,
  Sevilla, Carbillet, Carle, Cascone, Charton, Claudi, Costille, De~Caprio,
  Delboulbé, Dohlen, Engler, Fantinel, Feautrier, Fusco, Gigan, Girard, Giro,
  Gisler, Gluck, Gry, Hubin, Hugot, Jaquet, Kasper, Mignant, Llored, Madec,
  Magnard, Martinez, Maurel, Möller-Nilsson, Mouillet, Moulin, Origné,
  Pavlov, Perret, Petit, Pragt, Puget, Rabou, Ramos, Rickman, Rigal, Rochat,
  Roelfsema, Rousset, Roux, Salasnich, Sauvage, Sevin, Soenke, Stadler, Suarez,
  Wahhaj, Weber, \& Wildi}]{vigan_sphere_2020}
Vigan, A., Fontanive, C., Meyer, M., {et~al.} 2020, arXiv:2007.06573
  [astro-ph], arXiv: 2007.06573

\bibitem[{Virtanen {et~al.}(2020)Virtanen, Gommers, Oliphant, Haberland, Reddy,
  Cournapeau, Burovski, Peterson, Weckesser, Bright, van~der Walt, Brett,
  Wilson, Millman, Mayorov, Nelson, Jones, Kern, Larson, Carey, Polat, Feng,
  Moore, VanderPlas, Laxalde, Perktold, Cimrman, Henriksen, Quintero, Harris,
  Archibald, Ribeiro, Pedregosa, van Mulbregt, \& {SciPy 1. 0
  Contributors}}]{virtanen_scipy_2020}
Virtanen, P., Gommers, R., Oliphant, T.~E., {et~al.} 2020, Nature Methods, Vol.
  17, pp. 261-272, 17, 261

\bibitem[{Wahhaj {et~al.}(2015)Wahhaj, Cieza, Mawet, Yang, Canovas, de~Boer,
  Casassus, Ménard, Schreiber, Liu, Biller, Nielsen, \&
  Hayward}]{wahhaj_improving_2015}
Wahhaj, Z., Cieza, L.~A., Mawet, D., {et~al.} 2015, Astronomy \& Astrophysics,
  581, A24

\bibitem[{Wahhaj {et~al.}(2013)Wahhaj, Liu, Nielsen, Biller, Hayward, Close,
  Males, Skemer, Ftaclas, Chun, Thatte, Tecza, Shkolnik, Kuchner, Reid, Pino,
  Alencar, Gregorio-Hetem, Boss, Lin, \& Toomey}]{wahhaj_gemini_2013}
Wahhaj, Z., Liu, M.~C., Nielsen, E.~L., {et~al.} 2013, The Astrophysical
  Journal, 773, 179, publisher: IOP Publishing

\bibitem[{Wahhaj {et~al.}(2019)Wahhaj, Milli, Rodler, Girard, \&
  Vigan}]{wahhaj_sphere_2019}
Wahhaj, Z., Milli, J., Rodler, F., Girard, J., \& Vigan, A. 2019, SPHERE User
  Manual

\bibitem[{Walt {et~al.}(2014)Walt, Schönberger, Nunez-Iglesias, Boulogne,
  Warner, Yager, Gouillart, \& Yu}]{walt_scikit-image_2014}
Walt, S. v.~d., Schönberger, J.~L., Nunez-Iglesias, J., {et~al.} 2014, PeerJ,
  2, e453, publisher: PeerJ Inc.

\bibitem[{Wang {et~al.}(2018)Wang, Perrin, Savransky, Arriaga, Chilcote, Rosa,
  Millar-Blanchaer, Marois, Rameau, Wolff, Shapiro, Ruffio, Maire, Marchis,
  Graham, Macintosh, Ammons, Bailey, Barman, Bruzzone, Bulger, Cotten, Doyon,
  Duchêne, Fitzgerald, Follette, Goodsell, Greenbaum, Hibon, Hung, Ingraham,
  Kalas, Konopacky, Larkin, Marley, Metchev, Nielsen, Oppenheimer, Palmer,
  Patience, Poyneer, Pueyo, Rajan, Rantakyrö, Schneider, Sivaramakrishnan,
  Song, Soummer, Thomas, Wallace, Ward-Duong, \&
  Wiktorowicz}]{wang_automated_2018}
Wang, J.~J., Perrin, M.~D., Savransky, D., {et~al.} 2018, Journal of
  Astronomical Telescopes, Instruments, and Systems, 4, 018002, publisher:
  International Society for Optics and Photonics

\bibitem[{Zurlo {et~al.}(2016)Zurlo, Vigan, Galicher, Maire, Mesa, Gratton,
  Chauvin, Kasper, Moutou, Bonnefoy, Desidera, Abe, Apai, Baruffolo, Baudoz,
  Baudrand, Beuzit, Blancard, Boccaletti, Cantalloube, Carle, Cascone, Charton,
  Claudi, Costille, Caprio, Dohlen, Dominik, Fantinel, Feautrier, Feldt, Fusco,
  Gigan, Girard, Gisler, Gluck, Gry, Henning, Hugot, Janson, Jaquet, Lagrange,
  Langlois, Llored, Madec, Magnard, Martinez, Maurel, Mawet, Meyer, Milli,
  Moeller-Nilsson, Mouillet, Origné, Pavlov, Petit, Puget, Quanz, Rabou,
  Ramos, Rousset, Roux, Salasnich, Salter, Sauvage, Schmid, Soenke, Stadler,
  Suarez, Turatto, Udry, Vakili, Wahhaj, Wildi, \& Antichi}]{zurlo_first_2016}
Zurlo, A., Vigan, A., Galicher, R., {et~al.} 2016, Astronomy \& Astrophysics,
  587, A57, publisher: EDP Sciences

\end{thebibliography}
\end{document}